\documentclass[11pt]{article}
\usepackage{amssymb,amsmath,amsfonts}
\usepackage{graphicx}
\usepackage{graphics}
\usepackage{eepic,epsfig}

\makeatletter
\@addtoreset{equation}{section}

\makeatother

\begin{document}

\title{Surface polariton excitation and energy losses by a charged particle
in cylindrical waveguides}
\author{A. A. Saharian$^{1,2}$\thanks{%
E-mail: saharian@ysu.am}, L. Sh. Grigoryan$^{1}$, A. S. Kotanjyan$^{1}$,
\and H. F. Khachatryan$^{1}$ \vspace{0.3cm} \\
\textit{$^1$Institute of Applied Problems in Physics NAS RA, }\\
\textit{25 Hr. Nersessian Str., 0014 Yerevan, Armenia} \vspace{0.3cm}\\
\textit{$^2$ Department of Physics, Yerevan State University,}\\
\textit{1 Alex Manoogian Street, 0025 Yerevan, Armenia }}
\maketitle

\begin{abstract}
We investigate the emission of surface polaritons (SPs) by a charged
particle moving inside a dielectric cylinder parallel to its axis. It is
assumed that the cylinder is immersed into a homogeneous medium with
negative dielectric permittivity in the spectral range under consideration.
The SP modes are present for positive dielectric function of the cylinder
material. In order to find the electromagnetic fields corresponding to SPs,
the respective contributions to the components of the Green tensor are
separated. The expressions for scalar and vector potentials and for
electromagnetic field strengths are provided inside and outside the
cylinder. Those fields are expressed in terms of the SP eigenmodes of the
waveguide and we give detailed analysis for their properties. The SP energy
fluxes through the plane perpendicular to the cylinder axis are evaluated in
the interior and exterior media. The energy flux is directed towards the
charge motion inside the cylinder and towards the opposite direction in the
exterior region. The relativistic effects may essentially increase the
radiated energy. Important features of relativistic effects include the
possibility of essential increase of the radiated energy, the narrowing the
confinement region of the SP fields near the cylinder surface in the
exterior region, enlarging the frequency range for radiated SPs, and the
decrease of the cutoff factor for radiation at small wavelengths compared
with the waveguide radius. The general results are specified for the Drude
dispersion in the exterior medium. By using the Green tensor we also
evaluate the total energy losses of the charged particle for general case of
the interior and exterior dielectric functions. The corresponding results
are compared with those previously discussed in the literature. The
numerical data are presented in terms of scale invariant quantities that
allows to clarify the features of the SP radiation for different values of
the waveguide radius.
\end{abstract}

\bigskip


\bigskip

\section{Introduction}

The study of various aspects of generation and propagation of surface
polaritons (SPs, we use the term \textquotedblleft surface
polariton\textquotedblright\ and the related terminology in the sense
clarified in Ref. \cite{Welf91}) is an active field of research. They
present coupled excitations of the electromagnetic field and medium
polarization localized near the interface between two media. SPs are
generated in the spectral range where the real parts of dielectric
permittivities for neighboring materials have opposite signs. The
significant attention attracted by this type of surface waves is related to
their remarkable properties such as the enhancement of the electromagnetic
energy density, the possibility of concentrating the corresponding fields
beyond the diffraction limit for light waves, and the high sensitivity to
the electromagnetic characteristics of contacting materials. Further
specification of the type of SP is done based on the identification of
polarization mechanism in the negative-permittivity medium. An example of a
polarizable system is an electron gas in metals. The respective class of
surface waves is known as surface plasmon polaritons (SPPs, see, for
example, Refs. \cite{Welf91}-\cite{Stoc18}). Other negative-permittivity
materials supporting SP type waves are ionic crystals, semiconductors and
organic dielectrics. Another possibility is to use artificially constructed
materials referred to as metamaterials \cite{Marq08}. The spectral range for
SPs can be tuned by the choice of the negative-permittivity medium or by
using various mechanisms for changing the number density of charge carriers.
In particular, the metamaterials and doped semiconductors allow to excite
SPs in terahertz and microwave frequency ranges. Another possibility is to
use subwavelength microstructured interfaces. The unique properties of SPs
have resulted in wide range of applications that include biosensing, surface
imaging, data storage, solar cells, surface enhanced Raman spectroscopy,
nanophotonics, information processing systems, medicine and so forth.

Extensive applications of SPs in various fields of science and technology
motivate the development of efficient methods for excitation of that type of
waves with controllable characteristics. Available techniques for coupling
electromagnetic waves in free space to SPs, widely discussed in the
literature \cite{Agra82,Maie07,Enoc12,Han13}, include prism and grating
coupling, near-field scattering excitation and tight-focus excitation. SP
modes in waveguides can also be excited by using guided photonic modes from
another waveguide. Another source for SPs is provided by a beam of charged
particles passing through or near the interface between negative- and
positive-permittivity media. In fact, the first experimental signatures for
existence of SPs were obtained in measurements of electron beam energy
losses of aluminum and magnesium. High quality narrow electron beams of
scanning and transmission electron microscopes (SEM and TEM, respectively)
provide an excellent source in plasmonic devices with subnanometer
resolution (see \cite{Abaj10,Vess12}), essentially higher than the
resolutions realized with sources based on optical beams. The emission of
SPs is one of the channels for energy losses by charged particles. In
particular, motivated by applications in various research and technological
fields, the spectral distribution of total energy losses, including channels
of very different nature, has attracted a great deal of attention (for
reviews see, e.g., \cite{Abaj10,Riva00}). Different types of geometries for
separating surface have been discussed, including planar, spherical,
cylindrical geometries and structured interfaces such as gratings. The
corresponding length scales vary in rather wide intervals ranging from tens
of nanometers (e.g., for carbon nanotubes and fullerens) to millimeter and
centimeter sized objects like waveguides and resonators.

The energy losses of charged particles traveling parallel to the axis of a
cylindrical interface have been widely discussed in the literature (see,
e.g., \cite{Abaj10}, \cite{Bogd57}-\cite{Saha20} and references therein),
mainly within the framework of dielectric response theory. The carbon
nanotubes are among the interesting realizations of the corresponding setup.
The excitation of SPs in those structures by fast electrons have been
studied, for example, in \cite{Stoc99}-\cite{Mowb10}. The corresponding
results have important applications in the microscopy and spectroscopy of
materials and surfaces. The study of electron energy loss spectrum provides
a useful tool in investigations of features of both surface and bulk
collective excitations. The interaction between charged particles and
cylindrical interfaces is of utmost importance in the physics of particle
accelerators. In most of the existing literature dealing with that
interaction, the total energy losses have been considered which, in addition
to the radiation of SPs, include other channels as well. Our main concern in
the present paper is to investigate the energy fluxes of the radiated SPs
and their distribution in negative- and positive-permittivity media. The
total energy losses are separately considered as well by using the Green
tensor of the electromagnetic field from \cite{Grig95}. The corresponding
results are in agreement with those for the spectral density of the energy
loss probability previously considered in the literature.

The organization of the paper is as follows. In the next section, we
describe the problem setup and the components of the electromagnetic Green
tensor required for evaluation of fields are presented. The contributions in
the components coming from SPs are explicitly separated. In Section \ref%
{sec:AHE}, by using the expressions for the Green tensor components,
formulas are derived for the scalar and vector potentials and for the
electric and magnetic fields corresponding to radiated SPs. The properties
of the cylinder eigenmodes for SPs are discussed. The general expressions
are specified for the special case of axial motion. In Section \ref{sec:Flux}
the energy fluxes through the plane perpendicular to the cylinder axis are
investigated for SPs. The energy fluxes in the interior and exterior regions
are evaluated separately and the corresponding numerical results are
presented. Section \ref{sec:EnLoss} considers energy losses by a charged
particle for general case of interior and exterior dielectric
permittivities. A numerical example is provided for a vacuum cylindrical
hole inside a Drude like material. We then present our conclusions in
Section \ref{sec:Conc}.

\section{Problem setup and the contribution of surface polaritons to the
Green tensor}

\label{sec:GF}

\begin{figure}[tbph]
\begin{center}
\epsfig{figure=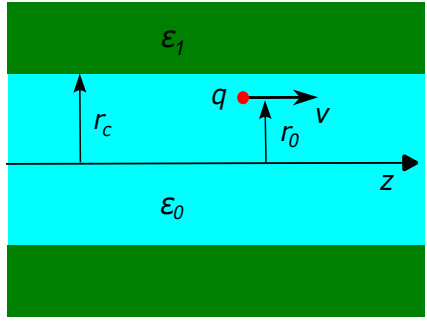,width=5.7cm,height=4.2cm}
\end{center}
\caption{The problem geometry and the notations.}
\label{fig1}
\end{figure}

Consider a point charge $q$ moving by a constant velocity $v$ parallel to
the axis of a cylinder with dielectric permittivity $\varepsilon _{0}$ and
with the radius $r_{c}$ (see Figure \ref{fig1}). The distance of the charge
trajectory from the axis will be denoted by $r_{0}<r_{c}$ and it will be
assumed that the cylinder is immersed into a homogeneous medium with
dielectric permittivity $\varepsilon _{1}$ (the magnetic permeabilities for
both of the cylinder and surrounding medium will be taken to be unit). In
accordance with the problem symmetry we will use cylindrical coordinates $%
(r,\phi ,z)$ with the axis $z$ along the axis of the cylinder. In the
present paper we are interested in the radiation from the charge in the form
of SPs.

Denoting by $x=(t,\mathbf{r})$ the spacetime point, the cylindrical
components of the vector potential $\mathbf{A}(x)$ are expressed in terms of
the electromagnetic field Green tensor $G_{il}(x,x^{\prime })$ as 
\begin{equation}
A_{i}(x)=-\frac{1}{2\pi ^{2}c}\int d^{4}x^{\prime
}\sum_{l=1}^{3}G_{il}(x,x^{\prime })j_{l}(x^{\prime }),  \label{Aix}
\end{equation}%
where $x^{\prime }=(t^{\prime },\mathbf{r}^{\prime })$ and the current
density is given by the expression 
\begin{equation}
j_{l}(x)=\delta _{3l}\frac{qv}{r}\delta (\mathbf{r}-\mathbf{r}_{0}(t)),
\label{jl}
\end{equation}%
where $\mathbf{r}_{0}(t)=(r=r_{0},\phi =0,z=vt)$ determines the location of
the charge.

For the Green tensor we use the partial Fourier expansion 
\begin{equation}
G_{il}(x,x^{\prime })=2\mathrm{Re}\left[ \sum_{n=-\infty }^{\infty
}\int_{-\infty }^{\infty }d\omega \int_{0}^{\infty }dk_{z}\,G_{il,n}(\omega
,k_{z},r,r^{\prime })e^{in\Delta \phi +ik_{z}\Delta z-i\omega \Delta t}%
\right] ,  \label{Gexp}
\end{equation}%
with $\Delta \phi =\phi -\phi ^{\prime }$, $\Delta z=z-z^{\prime }$, and $%
\Delta t=t-t^{\prime }$. Here the relation $G_{il,-n}(-\omega
,-k_{z},r,r^{\prime })=G_{il,n}^{\ast }(\omega ,k_{z},r,r^{\prime })$ is
used to transform the integral over the region $-\infty <k_{z}<+\infty $ to
the integral over $0\leq k_{z}<\infty $. In the problem under consideration
the Fourier components $G_{i3,n}(\omega ,k_{z},r,r^{\prime })$ are required,
evaluated for $r^{\prime }=r_{0}$. The corresponding expressions are given
in \cite{Grig95} (for applications of the Green tensor from \cite{Grig95} in
investigations of the Cherenkov and synchrotron radiations from a charge
rotating around/inside a dielectric cylinder see \cite{Saha05}-\cite{Saha19}
and references therein). They are presented in the form 
\begin{eqnarray}
G_{l3,n}(\omega ,k_{z},r,r_{0}) &=&-\sum_{p=\pm 1}\frac{p^{l-1}}{i^{l}}%
C_{n}^{(p)}J_{n+p}(\lambda _{0}r),\;l=1,2,  \notag \\
G_{33,n}(\omega ,k_{z},r,r_{0}) &=&\frac{\pi }{2i}\left[ J_{n}(\lambda
_{0}r_{<})H_{n}(\lambda _{0}r_{>})-\frac{V_{n}^{H}}{V_{n}^{J}}J_{n}(\lambda
_{0}r_{0})J_{n}(\lambda _{0}r)\right] ,  \label{Gi}
\end{eqnarray}%
for the region $r<r_{c}$ and by 
\begin{eqnarray}
G_{l3,n}(\omega ,k_{z},r,r_{0}) &=&-\sum_{p}\frac{p^{l-1}}{i^{l}}%
C_{n}^{(p)}H_{n+p}(\lambda _{1}r),\;l=1,2,  \notag \\
G_{33,n}(\omega ,k_{z},r,r_{0}) &=&\frac{J_{n}(\lambda _{0}r_{0})}{%
r_{c}V_{n}^{H}}H_{n}(\lambda _{1}r),  \label{Ge}
\end{eqnarray}%
in the region $r>r_{c}$. Here, $l=1,2,3$ correspond to the cylindrical
components $(r,\phi ,z)$, $J_{n}(y)$ and $H_{n}(y)=H_{n}^{(1)}(y)$ are the
Bessel and Hankel functions, $r_{<}=\mathrm{min}(r,r_{0})$, $r_{>}=\mathrm{%
max}(r,r_{0})$, and%
\begin{equation}
V_{n}^{F}=F_{n}(\lambda _{0}r_{c})\partial _{r_{c}}H_{n}(\lambda
_{1}r_{c})-H_{n}(\lambda _{1}r_{c})\partial _{r_{c}}F_{n}(\lambda _{0}r_{c}),
\label{WF}
\end{equation}%
with $F=J,H$, and%
\begin{equation}
\lambda _{j}^{2}=\omega ^{2}\varepsilon _{j}/c^{2}-k_{z}^{2},\;j=0,1.
\label{lambj}
\end{equation}%
The coefficients in (\ref{Gi}) and (\ref{Ge}) are defined by the relations%
\begin{equation}
C_{n}^{(p)}=\frac{k_{z}J_{n}(\lambda _{0}r_{0})H_{n}(\lambda _{1}r_{c})}{%
2r_{c}\alpha _{n}(\omega ,k_{z})V_{n}^{J}V_{n+p}^{J}}\left\{ 
\begin{array}{cc}
H_{n+p}(\lambda _{1}r_{c}), & r<r_{c} \\ 
J_{n+p}(\lambda _{0}r_{c}), & r>r_{c}%
\end{array}%
\right. ,  \label{Cnp}
\end{equation}%
where 
\begin{equation}
\alpha _{n}(\omega ,k_{z})=\frac{\varepsilon _{0}}{\varepsilon
_{1}-\varepsilon _{0}}+\frac{1}{2}\sum_{l=\pm 1}\left[ 1-\frac{\lambda _{1}}{%
\lambda _{0}}\frac{J_{n+l}(\lambda _{0}r_{c})H_{n}(\lambda _{1}r_{c})}{%
J_{n}(\lambda _{0}r_{c})H_{n+l}(\lambda _{1}r_{c})}\right] ^{-1}.
\label{alfn}
\end{equation}

The equation $\alpha _{n}(\omega ,k_{z})=0$ determines the electromagnetic
eigenmodes of the cylinder. It has solutions only under the condition $%
\lambda _{1}^{2}<0$. The corresponding fields exponentially decay in the
exterior medium. There are two types of the cylinder eigenmodes. For the
first one, corresponding to guiding modes, we have $\lambda _{0}^{2}>0$ and
the radial dependence of the fields inside the cylinder is given by the
Bessel functions $J_{n}(\lambda _{0}r)$ and $J_{n\pm 1}(\lambda _{0}r)$. For
the second type of the modes $\lambda _{0}^{2}<0$ and they correspond to
SPs. The corresponding radial dependence inside the cylinder is expressed in
terms of the modified Bessel functions $I_{n}(|\lambda _{0}|r)$ and $I_{n\pm
1}(|\lambda _{0}|r)$. Here we are interested in the radiation of SPs for
which $\lambda _{j}^{2}<0$. Introducing the modified Bessel functions, the
function determining the dispersion relation is presented as%
\begin{equation}
\alpha _{n}(\omega ,k_{z})=\frac{\varepsilon _{0}}{\varepsilon
_{1}-\varepsilon _{0}}+\frac{1}{2}\sum_{l=\pm 1}\left[ 1+\frac{|\lambda _{1}|%
}{|\lambda _{0}|}\frac{I_{n+l}(|\lambda _{0}|r_{c})K_{n}(|\lambda _{1}|r_{c})%
}{I_{n}(|\lambda _{0}|r_{c})K_{n+l}(|\lambda _{1}|r_{c})}\right] ^{-1}.
\label{alfSP}
\end{equation}%
From here, as a necessary condition for the presence of the roots for the
equation $\alpha _{n}(\omega ,k_{z})=0$ we get $\varepsilon _{1}/\varepsilon
_{0}<0$. This condition requiring opposite signs for the dielectric
permittivities of neighboring media is well known for planar interfaces. We
will denote by $k_{z}=k_{n}(\omega )$ the positive roots of the equation $%
\alpha _{n}(\omega ,k_{z})=0$. As it will be seen below, in the problem
under consideration for the radiated SPs one has $\omega =k_{z}v$. This
means that the eigenvalues of $k_{z}$ for the radiated SPs are determined by
the intersection of the dispersion curve $k_{n}(\omega )$ with the line $%
k_{z}=\omega /v$ in the $(\omega ,k_{z})$-plane. In the discussion below
those eigenvalues will be denoted by $k_{n,s}$, where $s$ enumerates the
roots for a given $n$. In the problem at hand the charge moves in the medium
with permittivity $\varepsilon _{0}$ and the most promising case to escape
the non-radiation energy losses and multiple scattering would be the motion
in an empty cylindrical hole with $\varepsilon _{0}=1$. Motivated by this,
we will specify the consideration for the case $\varepsilon
_{1}<0<\varepsilon _{0}$ assuming that the dielectric functions for both the
media are real. The total energy losses for general complex functions $%
\varepsilon _{0}(\omega )$ and $\varepsilon _{1}(\omega )$ will be discussed
below in Section \ref{sec:EnLoss}. With that choice, for a given value of $%
\beta _{0}=(v/c)\sqrt{\varepsilon _{0}}$, the roots with respect to 
\begin{equation}
u=k_{z}r_{c}=\omega r_{c}/v,  \label{u}
\end{equation}%
are functions of the combinations $\varepsilon _{1}/\varepsilon _{0}$ and $%
\beta _{0}$. Denoting those roots by $u_{n,s}=k_{n,s}r_{c}$, we get $%
u_{n,s}=u_{n,s}(\varepsilon _{1}/\varepsilon _{0},\beta _{0})$. In
particular, they do not depend on the cylinder radius.

In the integral over $k_{z}$ in (\ref{Gexp}) the integrand has poles at the
eigenmodes of the cylinder corresponding to $\alpha _{n}(\omega ,k_{z})=0$.
The specification of the integration contour is required near those poles.
In order to do that we introduce a small imaginary part for the dielectric
permittivity in the exterior medium writing it in the form $\varepsilon
_{1}=\varepsilon _{1}^{\prime }+i\varepsilon _{1}^{\prime \prime }$, where $%
\varepsilon _{1}^{\prime \prime }=\varepsilon _{1}^{\prime \prime }(\omega )=%
\mathrm{sgn}(\omega )|\varepsilon _{1}^{\prime \prime }(\omega )|$. For the
function $\alpha _{n}(\omega ,k_{z})$ this gives 
\begin{equation}
\alpha _{n}(\omega ,k_{z})\approx \alpha _{n}(\omega ,k_{z})|_{\varepsilon
_{1}=\varepsilon _{1}^{\prime }}+i\varepsilon _{1}^{\prime \prime }\partial
_{\varepsilon _{1}}\alpha _{n}(\omega ,k_{z})|_{\varepsilon _{1}=\varepsilon
_{1}^{\prime }}.  \label{alfnExp}
\end{equation}%
We have numerically checked that $\partial _{\varepsilon _{1}}\alpha
_{n}(\omega ,k_{z})|_{\varepsilon _{1}=\varepsilon _{1}^{\prime }}<0$ for
roots $k_{z}=k_{n,s}$. Taking the limit $\varepsilon _{1}^{\prime \prime
}\rightarrow 0$, it can be seen that near the poles, corresponding to the
radiated SPs, the factor $1/\alpha _{n}(\omega ,k_{z})$ in the integrand for
the Green tensor should be understood as $1/\alpha _{n}(\omega
,k_{z})\approx 1/[\alpha _{n}(\omega ,k_{z})-\mathrm{sgn}(\omega )i0]$. In
order to separate the respective contribution to the Green tensor we use the
relation%
\begin{equation}
\frac{1}{\alpha _{n}(\omega ,k_{z})-\mathrm{sgn}(\omega )i0}=\mathcal{P}%
\frac{1}{\alpha _{n}(\omega ,k_{z})}+\mathrm{sgn}(\omega )i\pi \delta
(\alpha _{n}(\omega ,k_{z})),  \label{alfnExp2}
\end{equation}%
where the symbol $\mathcal{P}$\ means that the corresponding part in the
integral should be understood in the sense of the principal value.

Let us denote by $G_{il}^{\mathrm{(P)}}(x,x^{\prime })$ the part of the
Green tensor coming from SPs. In addition, we will denote by $G_{il,n}^{%
\mathrm{(P)}}(\omega ,k_{z},r,r^{\prime })/\alpha _{n}(\omega ,k_{z})$ the
parts in the Fourier components containing the factor $1/\alpha _{n}(\omega
,k_{z})$. Those parts only contribute to the radiation of SPs. The part $%
G_{il}^{\mathrm{(P)}}(x,x^{\prime })$ is determined by the contribution of
the last term in (\ref{alfnExp2}) to the integral over $k_{z}$ in (\ref{Gexp}%
):%
\begin{equation}
G_{il}^{\mathrm{(P)}}(x,x^{\prime })=-2\pi \mathrm{Im}\left[ \sum_{n=-\infty
}^{\infty }\sum_{s}\int_{-\infty }^{\infty }d\omega \,\mathrm{sgn}(\omega )%
\frac{G_{il,n}^{\mathrm{(P)}}(\omega ,k_{z},r,r^{\prime })}{|\partial
_{k_{z}}\alpha _{n}(\omega ,k_{z})|}\left. e^{in\Delta \phi +ik_{z}\Delta
z-i\omega \Delta t}\right\vert _{k_{z}=k_{n,s}}\right] .  \label{GilP}
\end{equation}%
In particular, as it is seen from (\ref{Gi}) and (\ref{Ge}), $G_{33,n}^{%
\mathrm{(P)}}(\omega ,k_{z},r,r_{0})=0$ for both the exterior and interior
regions.

\section{Electromagnetic fields}

\label{sec:AHE}

Having the contribution of SPs to the Green tensor we can find the related
electromagnetic fields.

\subsection{Vector and scalar potentials}

Substituting the representation (\ref{GilP}) in (\ref{Aix}) and by using (%
\ref{jl}), for the vector potential corresponding to the radiated SPs we get%
\begin{equation}
A_{l}^{\mathrm{(P)}}(x)=\frac{2qv}{c}\mathrm{Im}\left[ \sum_{n=-\infty
}^{\infty }\sum_{s}\frac{G_{l3,n}^{\mathrm{(P)}}(k_{z}v,k_{z},r,r_{0})}{%
|\partial _{k_{z}}\alpha _{n}(\omega ,k_{z})|_{\omega =k_{z}v}}\left.
e^{in\phi +ik_{z}(z-vt)}\right\vert _{k_{z}=k_{n,s}}\right] .  \label{AlP}
\end{equation}%
The expressions for $G_{i3,n}^{\mathrm{(P)}}(k_{z}v,k_{z},r,r_{0})$ are
obtained from (\ref{Gi}) and (\ref{Ge}):%
\begin{equation}
G_{l3,n}^{\mathrm{(P)}}(k_{z}v,k_{z},r,r_{0})=-\frac{i^{-l}}{2r_{c}}%
\sum_{p}p^{l-1}I_{n}(\gamma _{0}ur_{0}/r_{c})K_{n}(\gamma _{1}u)\frac{%
R_{n+p}(u,r/r_{c})}{uW_{n}^{I}W_{n+p}^{I}},  \label{Gl3P}
\end{equation}%
for $l=1,2$, and $G_{33,n}^{\mathrm{(P)}}(k_{z}v,k_{z},r,r_{0})=0$. In (\ref%
{Gl3P}) we use the notations (\ref{u}) and 
\begin{equation}
R_{n}(u,r/r_{c})=\left\{ 
\begin{array}{cc}
K_{n}(\gamma _{1}u)I_{n}(\gamma _{0}ur/r_{c}), & r<r_{c}, \\ 
I_{n}(\gamma _{0}u)K_{n}(\gamma _{1}ur/r_{c}), & r>r_{c}.%
\end{array}%
\right. ,  \label{Rn}
\end{equation}%
with%
\begin{equation}
\gamma _{j}=\sqrt{1-\beta ^{2}\varepsilon _{j}},\;\beta =v/c,  \label{gamj}
\end{equation}%
and%
\begin{equation}
W_{n}^{I}=-\gamma _{1}I_{n}(\gamma _{0}u)K_{n+1}(\gamma _{1}u)-\gamma
_{0}I_{n+1}(\gamma _{0}u)K_{n}(\gamma _{1}u).  \label{WnF}
\end{equation}%
For SPs under consideration $\lambda _{j}=i|k_{z}|\gamma _{j}$ and the
function $\alpha _{n}(\omega ,k_{z})=\alpha _{n}(k_{z}v,k_{z})\equiv \alpha
_{n}(u)$ for those modes is written in the form%
\begin{equation}
\alpha _{n}(u)=\frac{\varepsilon _{0}}{\varepsilon _{1}-\varepsilon _{0}}+%
\frac{1}{2}\sum_{l=\pm 1}\left[ 1+\frac{\gamma _{1}}{\gamma _{0}}\frac{%
I_{n+l}(\gamma _{0}u)K_{n}(\gamma _{1}u)}{I_{n}(\gamma _{0}u)K_{n+l}(\gamma
_{1}u)}\right] ^{-1}.  \label{SPmodes}
\end{equation}%
Note that we have assumed $\varepsilon _{1}<0<\varepsilon _{0}$ and, hence, $%
\gamma _{0}<1<\gamma _{1}$.

The insertion of (\ref{Gl3P}) in (\ref{AlP}) gives%
\begin{equation}
A_{l}^{\mathrm{(P)}}(x)=\frac{2qv}{cr_{c}}\sideset{}{'}{\sum}_{n=0}^{\infty
}\sum_{s}Q_{n}(u)\sum_{p=\pm 1}\frac{R_{n+p}(u,r/r_{c})}{p^{l-1}uW_{n+p}^{I}}%
\cos \left( u\xi /r_{c}\right) \sin \left( l\pi /2-n\phi \right)
|_{u=u_{n,s}},  \label{AlP2}
\end{equation}%
for $l=1,2$, $A_{3}^{\mathrm{(P)}}(x)=0$, and the prime on the summation
sign means that the term $n=0$ should be taken with an additional
coefficient 1/2. Here 
\begin{equation}
u_{n,s}=k_{n,s}r_{c},\;\xi =vt-z,  \label{ksi}
\end{equation}%
and 
\begin{equation}
Q_{n}(u)=\frac{K_{n}(\gamma _{1}u)}{W_{n}^{I}\bar{\alpha}_{n}(u)}%
I_{n}(\gamma _{0}ur_{0}/r_{c}).  \label{Qn}
\end{equation}%
In (\ref{Qn}) and in what follows we use the notation%
\begin{equation}
\bar{\alpha}_{n}(u)=|\partial _{u}\alpha _{n}(\omega ,u/r_{c})|_{\omega
=uv/r_{c}}.  \label{alfnb}
\end{equation}%
Note that one has 
\begin{equation}
W_{n+p}^{I}=-\gamma _{1}I_{n+p}(\gamma _{0}u)K_{n}(\gamma _{1}u)-\gamma
_{0}I_{n}(\gamma _{0}u)K_{n+p}(\gamma _{1}u),  \label{WnF2}
\end{equation}%
and the expression (\ref{SPmodes}) for the function $\alpha _{n}(u)$ is
rewritten in the form%
\begin{equation}
\alpha _{n}(u)=\frac{\varepsilon _{0}}{\varepsilon _{1}-\varepsilon _{0}}-%
\frac{1}{2}\gamma _{0}I_{n}(\gamma _{0}u)\sum_{p=\pm 1}\frac{K_{n+p}(\gamma
_{1}u)}{W_{n+p}^{I}}.  \label{alf2}
\end{equation}%
As expected, the vector potential is continuous at the cylinder surface $%
r=r_{c}$. It is important to note that, in general, the function $\bar{\alpha%
}_{n}(u)$ differs from the derivative $\alpha _{n}^{\prime }(u)$ of the
function (\ref{alf2}). By using (\ref{WnF2}), The equation $\alpha
_{n}(k_{z}v,k_{z})=0$ for SP eigenmodes $u=u_{n,s}$ is reduced to (see also 
\cite{Ashl74,Khos91} and Ref. \cite{Jack99} for the corresponding guiding
modes) 
\begin{equation}
\left[ \gamma _{1}\frac{I_{n}^{\prime }(\gamma _{0}u)}{I_{n}(\gamma _{0}u)}%
-\gamma _{0}\frac{K_{n}^{\prime }(\gamma _{1}u)}{K_{n}(\gamma _{1}u)}\right] %
\left[ \varepsilon _{0}\gamma _{1}\frac{I_{n}^{\prime }\left( \gamma
_{0}u\right) }{I_{n}\left( \gamma _{0}u\right) }-\varepsilon _{1}\gamma _{0}%
\frac{K_{n}^{\prime }\left( \gamma _{1}u\right) }{K_{n}\left( \gamma
_{1}u\right) }\right] =\left( \frac{n\beta }{u}\frac{\varepsilon
_{0}-\varepsilon _{1}}{\gamma _{0}\gamma _{1}}\right) ^{2},  \label{EigEq}
\end{equation}%
where the prime stands for the derivative of functions with respect to the
argument.

The spectral component of the scalar potential $\varphi ^{\mathrm{(P)}}(x)$,
denoted here as $\varphi _{\omega }^{\mathrm{(P)}}$, with $\omega
=u_{n,s}v/r_{c}$ and $\varphi _{\omega }^{\mathrm{(P)}}=\int_{-\infty
}^{\infty }dt\,\varphi ^{\mathrm{(P)}}(x)e^{i\omega t}/2\pi $, is found by
using the relation $\varphi _{\omega }^{\mathrm{(P)}}=-\left( ic/\omega
\varepsilon \right) \nabla \cdot \mathbf{A}_{\omega }^{\mathrm{(P)}}$ in
separate regions $r>r_{c}$ and $r<r_{c}$. This gives%
\begin{equation}
\varphi ^{\mathrm{(P)}}(x)=-\frac{2q}{r_{c}}\sideset{}{'}{\sum}%
_{n=0}^{\infty }\sum_{s}Q_{n}(u)\sum_{p=\pm 1}\frac{R_{p,n}^{\mathrm{(e)}%
}(u,r/r_{c})}{uW_{n+p}^{I}}\sin \left( u\xi /r_{c}\right) \cos \left( n\phi
\right) |_{u=u_{n,s}},  \label{phiP}
\end{equation}%
with the notation%
\begin{equation}
R_{p,n}^{\mathrm{(e)}}(u,r/r_{c})=\left\{ 
\begin{array}{cc}
\frac{\gamma _{0}}{\varepsilon _{0}}K_{n+p}(\gamma _{1}u)I_{n}(\gamma
_{0}ur/r_{c}), & r<r_{c} \\ 
-\frac{\gamma _{1}}{\varepsilon _{1}}I_{n+p}(\gamma _{0}u)K_{n}(\gamma
_{1}ur/r_{c}), & r>r_{c}%
\end{array}%
\right. .  \label{Rpne}
\end{equation}%
We can check that the spectral components of the scalar potential obey the
boundary condition 
\begin{equation}
\varepsilon _{0}\partial _{r}\varphi _{\omega }^{\mathrm{(P)}%
}|_{r=r_{c}-0}-\varepsilon _{1}\partial _{r}\varphi _{\omega }^{\mathrm{(P)}%
}|_{r=r_{c}+0}=i\left( \varepsilon _{0}-\varepsilon _{1}\right) \frac{\omega 
}{c}A_{\omega 1}^{\mathrm{(P)}}|_{r=r_{c}},  \label{phibc}
\end{equation}%
where the expression for the spectral component of the vector potential on
the cylinder surface directly follows from (\ref{AlP2}).

It is important to mention that the potentials $\varphi ^{\mathrm{(P)}}(x)$
and $\mathbf{A}^{\mathrm{(P)}}(x)$ are the parts of the total fields
corresponding to radiated SPs. The total fields can be found
by using the relation (\ref{Aix}) and the components (\ref{Gi}) and (\ref{Ge}%
) for the electromagnetic field Green tensor. The part in the fields coming
from the first term in the square brackets of the expression (\ref{Gi}) for
the component $G_{33,n}(\omega ,k_{z},r,r_{0})$ corresponds to the field
generated by a charged particle moving in a homogeneous medium with
dielectric permittivity $\varepsilon _{0}$. Denoting by $\mathbf{A}%
^{(0)}(x)=(0,0,A_{3}^{(0)}(x))$ the corresponding vector potential, we get 
\begin{equation}
A_{3}^{(0)}(x)=-\frac{qv}{2ic}\int_{-\infty }^{\infty
}dk_{z}e^{ik_{z}(z-vt)}\sum_{n=-\infty }^{\infty }e^{in\phi }J_{n}(\lambda
_{0}r_{<})H_{n}(\lambda _{0}r_{>}).  \label{vecpot2}
\end{equation}%
The series over $n$ in this expression is summed by using the addition
theorem for the cylinder functions (see, for example, \cite{Abra72}). By
taking into account that $\lambda _{0}=ik_{z}\gamma _{0}$, the sum of the
series is equal to $2K_{0}(k_{z}\gamma _{0}r_{\perp })/(\pi i)$, where $%
r_{\perp }=\sqrt{r^{2}+r_{0}^{2}-2rr_{0}\cos \phi }$ is the distance of the
observation point $(r,\phi ,z)$ from the trajectory of the charge. For the
spectral component of the vector potential, $A_{\omega 3}^{(0)}$, this gives 
$A_{\omega 3}^{(0)}=qe^{i\frac{\omega }{v}z}K_{0}(\frac{\omega }{v}\gamma
_{0}b)/(\pi c)$. The corresponding scalar potential is obtained from the
relation $\varphi _{\omega }^{(0)}=-\left( ic/\omega \varepsilon _{0}\right)
\partial _{z}A_{\omega 3}^{(0)}$ and one gets $\varphi _{\omega
}^{(0)}=A_{\omega 3}^{(0)}/(\beta \varepsilon _{0})$. The spectral component
of the electric field strength is found by using the formula $\mathbf{E}%
_{\omega }^{(0)}=i\omega \mathbf{A}_{\omega }^{(0)}/c-\mathbf{\nabla }%
\,\varphi _{\omega }^{(0)}$. In particular, for the $z$-projection we find 
\begin{equation}
E_{\omega 3}^{(0)}=-\frac{iq\omega }{\pi v^{2}}\left( \,\frac{1}{\varepsilon
_{0}}-\beta ^{2}\right) e^{i\frac{\omega }{v}z}K_{0}(\frac{\omega }{v}\gamma
_{0}r_{\perp }).  \label{E03}
\end{equation}%
For $z=0$ this result differs from the corresponding expression given in Ref.
\cite{Jack99} (see Eq. (13.32)) by an additional coefficient $1/\sqrt{2\pi }$
which is related to different coefficients in the definition of the Fourier transformation.

\subsection{Magnetic and electric fields}

The cylindrical components of the magnetic field $\mathbf{H}$ are found from
the relation $\mathbf{H}=\nabla \times \mathbf{A}$. Denoting the part
related to SPs by $\mathbf{H}^{\mathrm{(P)}}$ and by making use of the
recurrence relations for the modified Bessel functions one obtains%
\begin{eqnarray}
H_{l}^{\mathrm{(P)}}(x) &=&\sum_{n=0}^{\infty }\sum_{s}H_{l,n}^{\mathrm{(P)}%
}\left( u_{n,s}\right) \sin \left( u_{n,s}\xi /r_{c}\right) \cos (\pi
l/2-n\phi ),  \notag \\
H_{3}^{\mathrm{(P)}}(x) &=&\sum_{n=0}^{\infty }\sum_{s}H_{3,n}^{\mathrm{(P)}%
}\left( u_{n,s}\right) \cos \left( u_{n,s}\xi /r_{c}\right) \sin \left(
n\phi \right) ,  \label{H3P}
\end{eqnarray}%
where%
\begin{eqnarray}
H_{l,n}^{\mathrm{(P)}}\left( u\right) &=&-2\delta _{n}\frac{q\beta }{%
r_{c}^{2}}Q_{n}(u)\sum_{p=\pm 1}\frac{R_{n+p}(u,r/r_{c})}{p^{l}W_{n+p}^{I}},
\notag \\
H_{3,n}^{\mathrm{(P)}}\left( u\right) &=&2\delta _{n}\frac{q\beta }{r_{c}^{2}%
}Q_{n}(u)\sum_{p=\pm 1}\frac{R_{p,n}^{\mathrm{(m)}}(u,r/r_{c})}{pW_{n+p}^{I}}%
,  \label{H3Pn}
\end{eqnarray}%
and $\delta _{0}=1/2$, $\delta _{n}=1$ for $n>1$. In (\ref{H3Pn}) we have
defined the function%
\begin{equation}
R_{p,n}^{\mathrm{(m)}}(u,r/r_{c})=\left\{ 
\begin{array}{cc}
\gamma _{0}K_{n+p}(\gamma _{1}u)I_{n}(\gamma _{0}ur/r_{c}), & r<r_{c} \\ 
-\gamma _{1}I_{n+p}(\gamma _{0}u)K_{n}(\gamma _{1}ur/r_{c}), & r>r_{c}%
\end{array}%
\right. ,  \label{Rnm}
\end{equation}%
for the exterior and interior regions. It can be checked that the magnetic
field is continuous on the cylinder surface.

From the relation $\mathbf{E}_{\omega }=i\omega \mathbf{A}_{\omega }/c-%
\mathbf{\nabla }\,\varphi _{\omega }$ we find the spectral components of the
electric field for radiated SPs. The corresponding Fourier expansions have
the form%
\begin{eqnarray}
E_{l}^{\mathrm{(P)}}(x) &=&\sum_{n=0}^{\infty }\sum_{s}E_{l,n}^{\mathrm{(P)}%
}\left( u_{n,s}\right) \sin \left( u_{n,s}\xi /r_{c}\right) \sin \left( \pi
l/2-n\phi \right) ,  \notag \\
E_{3}^{\mathrm{(P)}}(x) &=&\sum_{n=0}^{\infty }\sum_{s}E_{3,n}^{\mathrm{(P)}%
}\left( u_{n,s}\right) \cos \left( u_{n,s}\xi /r_{c}\right) \cos (n\phi ),
\label{ElP}
\end{eqnarray}%
where $l=1,2$. The Fourier components are expressed as%
\begin{eqnarray}
E_{l,n}^{\mathrm{(P)}}\left( u\right) &=&\delta _{n}\frac{q}{r_{c}^{2}}%
Q_{n}(u)\sum_{p,p^{\prime }=\pm 1}\frac{1+pp^{\prime }\beta _{0}^{2}}{p^{l-1}%
}\frac{K_{n+p^{\prime }}(\gamma _{1}u)}{\varepsilon _{0}W_{n+p^{\prime }}^{I}%
}I_{n+p}(\gamma _{0}ur/r_{c}),\;r<r_{c},  \notag \\
E_{l,n}^{\mathrm{(P)}}\left( u\right) &=&\delta _{n}\frac{q}{r_{c}^{2}}%
Q_{n}(u)\sum_{p,p^{\prime }=\pm 1}\frac{1+pp^{\prime }\beta ^{2}\varepsilon
_{1}}{p^{l-1}}\frac{I_{n+p^{\prime }}(\gamma _{0}u)}{\varepsilon
_{1}W_{n+p^{\prime }}^{I}}K_{n+p}(\gamma _{1}ur/r_{c}),\;r>r_{c},
\label{ElPn}
\end{eqnarray}%
and%
\begin{equation}
E_{3,n}^{\mathrm{(P)}}\left( u\right) =-\frac{2\delta _{n}q}{r_{c}^{2}}%
\sum_{p=\pm 1}\frac{Q_{n}(u)}{W_{n+p}^{I}}R_{p,n}^{\mathrm{(e)}}(u,r/r_{c}).
\label{E3Pn}
\end{equation}%
It can be checked that the components $E_{2}^{\mathrm{(P)}}(x)$ and $E_{3}^{%
\mathrm{(P)}}(x)$ for the electric field and the component $D_{1}^{\mathrm{%
(P)}}(x)$ for the displacement vector are continuous on the boundary $%
r=r_{c} $. The expressions for separate components can be further simplified
by using the relations 
\begin{eqnarray}
\sum_{p=\pm 1}\frac{K_{n+p}(\gamma _{1}u)}{W_{n+p}^{I}} &=&\frac{%
2\varepsilon _{0}/(\varepsilon _{1}-\varepsilon _{0})}{\gamma
_{0}I_{n}(\gamma _{0}u)},  \notag \\
\sum_{p=\pm 1}\frac{I_{n+p}(\gamma _{0}u)}{W_{n+p}^{I}} &=&\frac{%
2\varepsilon _{1}/\left( \varepsilon _{0}-\varepsilon _{1}\right) }{\gamma
_{1}K_{n}(\gamma _{1}u)},  \label{RelK}
\end{eqnarray}%
which directly follows from $\alpha _{n}(u)=0$ in combination with (\ref%
{alf2}). Note that both the electric and magnetic fields, in addition to the
transversal components, have longitudinal components.

The expressions for the fields are further simplified on the axis of the
cylinder. The cylindrical coordinates are degenerate for $r=0$ and, in order
to take the limit $r\rightarrow 0$, we first transform the fields to
Cartesian coordinates $(x,y,z)$, where the $x$-axis corresponds to $\phi =0$
and for the location of the charge one has $(x,y,z)=(r_{0},0,vt)$. In this
way, for the nonzero Cartesian components of the electric and magnetic
fields on the cylinder axis, $r=0$, one finds%
\begin{eqnarray}
H_{y}^{\mathrm{(P)}} &=&\frac{2q\beta }{r_{c}^{2}}\sum_{s}\frac{Q_{1}(u)}{%
W_{0}^{I}}K_{0}(\gamma _{1}u)\sin \left( u\xi /r_{c}\right) |_{u=u_{1,s}}, 
\notag \\
E_{x}^{\mathrm{(P)}} &=&\frac{2q}{r_{c}^{2}}\sum_{s}Q_{1}(u)\left[ \frac{%
\beta ^{2}}{W_{0}^{I}}K_{0}(\gamma _{1}u)-\frac{\gamma _{0}/(\varepsilon
_{0}-\varepsilon _{1})}{I_{1}(\gamma _{0}u)}\right] \sin \left( u\xi
/r_{c}\right) |_{u=u_{1,s}},  \notag \\
E_{z}^{\mathrm{(P)}} &=&-\frac{q}{r_{c}^{2}}\sum_{s}\frac{\gamma _{0}Q_{0}(u)%
}{\varepsilon _{0}W_{1}^{I}}K_{1}(\gamma _{1}u)\cos \left( u\xi
/r_{c}\right) |_{u=u_{0,s}}.  \label{Ezaxis}
\end{eqnarray}%
Hence, on the axis the electric and magnetic fields are orthogonal and the
magnetic field is transversal

We recall that for given $\beta _{0}$ and $\varepsilon _{1}/\varepsilon _{0}$
the eigenmodes $u_{n,s}$ do not depend on the cylinder radius $r_{c}$. The
eigenvalues for the SP wavelength are expressed as $\lambda _{\mathrm{sp}%
}=\lambda _{n,s}=2\pi r_{c}/u_{n,s}$. The dependence of the Fourier
components $E_{l,n}^{\mathrm{(P)}}\left( u_{n,s}\right) $ and $H_{l,n}^{%
\mathrm{(P)}}\left( u_{n,s}\right) $ on the cylinder radius appears in the
form of the coefficient $q/r_{c}^{2}$ and in the form of the ratios $%
r_{0}/r_{c}$ and $r/r_{c}$ in the arguments of the modified Bessel
functions. Note that the dependence on the impact parameter $r_{0}$ comes
from the function $I_{n}(\gamma _{0}ur_{0}/r_{c})$ in the definition (\ref%
{Qn}) for $Q_{n}(u)$. Hence, for fixed values of the other parameters the
absolute values $|E_{l,n}^{\mathrm{(P)}}\left( u_{n,s}\right) |$ and $%
|H_{l,n}^{\mathrm{(P)}}\left( u_{n,s}\right) |$ monotonically increase with
increasing $r_{0}<r_{c}$. The fields in the negative-permittivity medium
depend on the radial coordinate through the functions $K_{n}(\gamma
_{1}u_{n,s}r/r_{c})$, $K_{n\pm 1}(\gamma _{1}u_{n,s}r/r_{c})$ and, hence,
they are exponentially suppressed at distances $r\gg \lambda _{n,s}/\left(
2\pi \gamma _{1}\right) $. For a given wavelength, the SP fields in the
negative-permittivity medium are mainly localized near the cylinder surface
within the region of the thickness $\lesssim \lambda _{n,s}/\left( 2\pi
\gamma _{1}\right) $. In the problem at hand $\gamma _{1}>1$ and the
localization radius can be essentially smaller than the wavelength. Note
that the parameter $\gamma _{1}$ increases with increasing $\beta $ and the
localization near the cylinder surface in the exterior region is stronger
for relativistic electrons (for the discussion of relativistic effects in
radiation of SPs see also \cite{Abaj10}).

\subsection{Properties of the modes}

The electromagnetic fields for SPs are expressed in terms of the
corresponding eigenmodes of the cylinder. In this subsection the properties
of those modes are discussed. We can consider the equation $\alpha _{n}(u)=0$%
, with $\alpha _{n}(u)$ from (\ref{SPmodes}), as an equation determining the
ratio $\varepsilon _{1}/\varepsilon _{0}$ for given values $u$ and $\beta
_{0}$: $\varepsilon _{1}/\varepsilon _{0}=f_{n}(u,\beta _{0})$. Let us
clarify the asymptotic properties of the roots interpreted in that way. For
large values of $u$, assuming that $\gamma _{j}u\gg 1$, by using the
asymptotic expressions for the modified Bessel functions (see, for example, 
\cite{Abra72}) in (\ref{SPmodes}) we get%
\begin{equation}
\alpha _{n}(u)\approx \frac{\varepsilon _{0}}{\varepsilon _{1}-\varepsilon
_{0}}+\left( 1+\frac{\gamma _{1}}{\gamma _{0}}\right) ^{-1}\left( 1+\frac{1}{%
2\gamma _{0}u}\right) .  \label{alfnasu}
\end{equation}%
From here it follows that in the limit $u\rightarrow \infty $ the roots of
the equation $\alpha _{n}(u)=0$ with respect to $\varepsilon
_{1}/\varepsilon _{0}$ tend to the limiting value $1/(\beta _{0}^{2}-1)$.
Note that to the leading order one has 
\begin{equation}
\gamma _{1}/\gamma _{0}\approx -\varepsilon _{1}/\varepsilon _{0}.
\label{gamas}
\end{equation}%
This is an exact relation in the limit of a planar boundary. By taking into
account that $u=2\pi r_{c}/\lambda _{\mathrm{sp}}$, we see that the large
values for $u$ correspond to wavelengths much smaller than the cylinder
radius. We could expect that in this limit the curvature effects of the
separating boundary will be weak.

In the opposite limit $u\ll 1$ and for the modes with $n=0$ one finds%
\begin{equation}
\alpha _{0}(u)\approx \frac{\varepsilon _{1}}{\varepsilon _{1}-\varepsilon
_{0}}+\frac{1}{4}u^{2}\gamma _{1}^{2}\ln (u\gamma _{1}).  \label{alhnasu2}
\end{equation}%
This shows that for those modes the roots with respect to $\varepsilon
_{1}/\varepsilon _{0}$ tend to zero. For $n\geq 1$ and $u\ll 1$ the
asymptotic expressions for the function $\alpha _{n}(u)$ have the form%
\begin{eqnarray}
\alpha _{1}(u) &\approx &\frac{1+\varepsilon _{1}/\varepsilon _{0}}{2\left(
\varepsilon _{1}/\varepsilon _{0}-1\right) }-\frac{\gamma _{1}^{2}u^{2}}{16}-%
\frac{1}{4}\gamma _{0}^{2}u^{2}\ln (\gamma _{1}u),  \notag \\
\alpha _{n}(u) &\approx &\frac{1+\varepsilon _{1}/\varepsilon _{0}}{2\left(
\varepsilon _{1}/\varepsilon _{0}-1\right) }+u^{2}\frac{2+\left[ \left(
n-1\right) \varepsilon _{1}-\left( n+1\right) \varepsilon _{0}\right] \beta
^{2}}{8n\left( n^{2}-1\right) },  \label{alfnasu3}
\end{eqnarray}%
with $n\geq 2$ in the second line. From here we conclude that for $n\geq 1$
the roots $\varepsilon _{1}/\varepsilon _{0}$ tend to $-1$.

Now let us consider the properties of SP modes for large values of $n$. By
using the uniform asymptotic expansions of the modified Bessel function for
large values of the order \cite{Abra72}, to the leading order we get 
\begin{equation}
\alpha _{n}(u)\approx \frac{\varepsilon _{0}}{\varepsilon _{1}-\varepsilon
_{0}}+\frac{1}{2}\sum_{l=\pm 1}\left( 1+\frac{\gamma _{1}^{2}}{\gamma
_{0}^{2}}\frac{\sqrt{1+u^{2}\gamma _{0}^{2}/n^{2}}-l}{\sqrt{1+u^{2}\gamma
_{1}^{2}/n^{2}}+l}\right) ^{-1}.  \label{alfln}
\end{equation}%
First let us consider the possibility of the modes with $u\gamma _{j}\ll n$
when the leading order term is reduced to%
\begin{equation}
\alpha _{n}(u)\approx \frac{\varepsilon _{1}+\varepsilon _{0}}{\varepsilon
_{1}-\varepsilon _{0}}+\mathcal{O}\left( 1/n^{2}\right) .  \label{alfln1}
\end{equation}%
From here it follows that for large $n$ the corresponding modes are present
if the ratio $\varepsilon _{1}/\varepsilon _{0}$ is sufficiently close to $%
-1 $: $\varepsilon _{1}/\varepsilon _{0}\approx -1+\mathcal{O}\left(
1/n^{2}\right) $. For the modes with $u\gamma _{j}$ of the order $n\gg 1$,
solving the equation $\alpha _{n}(u)=0$, with $\alpha _{n}(u)$ from (\ref%
{alfln}), we get%
\begin{equation}
u_{n,s}\approx n\left( \frac{\beta _{0}^{2}\varepsilon _{1}}{\varepsilon
_{0}+\varepsilon _{1}}-1\right) ^{-1/2}.  \label{unsln}
\end{equation}%
In the region under consideration the neighboring roots with respect to $n$
are approximately equidistant.

\begin{figure}[tbph]
\begin{center}
\begin{tabular}{cc}
\epsfig{figure=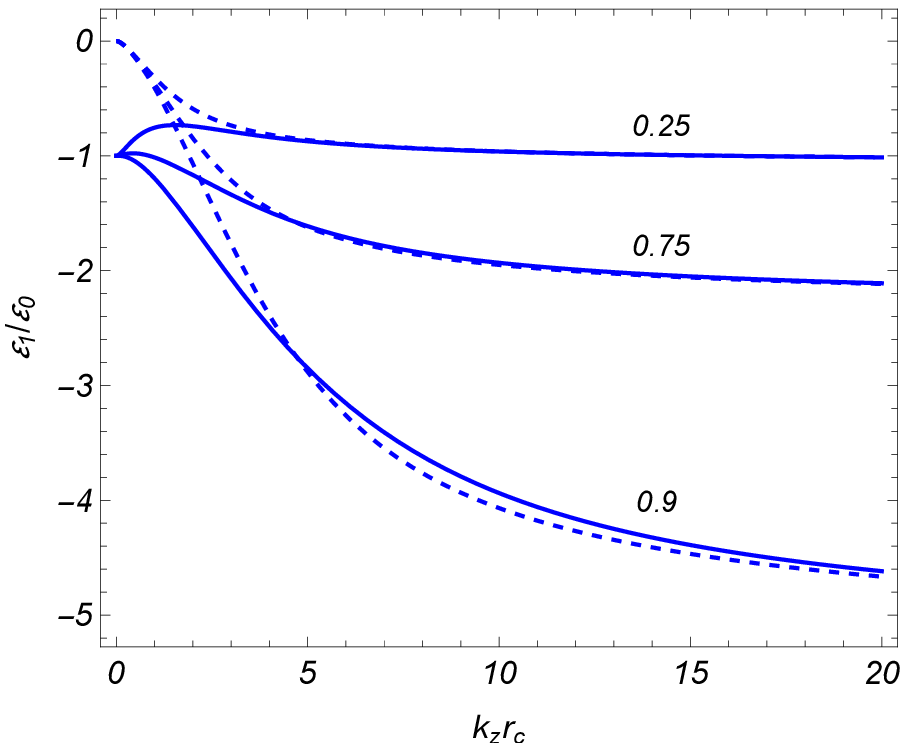,width=7.5cm,height=6.cm} & \quad %
\epsfig{figure=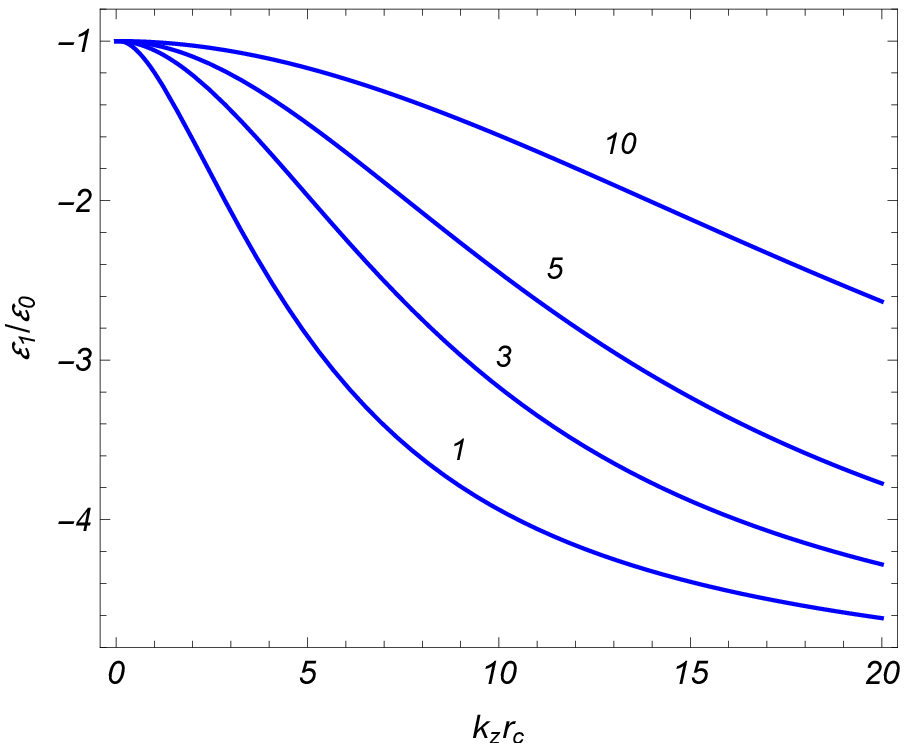,width=7.5cm,height=6.cm}%
\end{tabular}%
\end{center}
\caption{The roots of the eignemode equation for SPs with respect to the
ratio $\protect\varepsilon _{1}/\protect\varepsilon _{0}$ as functions of $%
u=k_{z}r_{c}$. The dashed and full curves on the left panel correspond to
the modes with $n=0$ and $n=1$, respectively, and the numbers near the
curves are the values of $\protect\beta _{0}$. The right panel is plotted
for $\protect\beta _{0}=0.9$ and for different values of $n$ (the numbers
near the curves).}
\label{fig2}
\end{figure}

In Figure \ref{fig2} we have presented the distribution of the roots for the
equation $\alpha _{n}(u)=0$ with respect to $\varepsilon _{1}/\varepsilon
_{0}$ as functions of $u$. On the left panel the graphs are plotted for $n=0$
and $n=1$ (the dashed and full curves respectively) and for fixed values of $%
\beta _{0}$ (the numbers near the curves). The right panel presents the
graphs for different values of $n$ (the numbers near the curves) and for $%
\beta _{0}=0.9$. The numerical data confirm the features clarified by the
asymptotic analysis: $\varepsilon _{1}/\varepsilon _{0}$ tends to $-1+\delta
_{0n}$ in the limit $u\rightarrow 0$ and $\varepsilon _{1}/\varepsilon
_{0}\rightarrow 1/(\beta _{0}^{2}-1)$ for $u\rightarrow \infty $.

We have considered the general properties of SP modes without fixing the
dispersion law for the dielectric functions $\varepsilon _{j}(\omega )$.
Figure \ref{fig2} displays the function $\varepsilon _{1}/\varepsilon
_{0}=f_{n}(u,\beta _{0})$ for different values of $\beta _{0}$ and $n$.
Given the dielectric functions $\varepsilon _{j}(\omega )=\varepsilon
_{j}(uv/r_{c})$, the eigenvalues for $u$ are determined by the intersections
of the graphs for the functions $\varepsilon _{1}(uv/r_{c})/\varepsilon
_{0}(uv/r_{c})$ and $f_{n}(u,\beta _{0})$. As an example we will consider
the Drude model for the exterior medium assuming that the dispersion of the
material of the cylinder in the frequency range under consideration is weak.
The simplest example would be the motion of the charge in an empty
cylindrical hole with $\varepsilon _{0}=1$. Denoting by $\omega _{p}$ and $%
\eta $ the plasma frequency and the characteristic collision frequency, the
function $\varepsilon _{1}=\varepsilon _{1}(\omega )$ is expressed as 
\begin{equation}
\varepsilon _{1}(\omega )=1-\frac{\omega _{p}^{2}}{\omega ^{2}+i\eta \omega }%
.  \label{eps1}
\end{equation}%
In accordance with the assumption made above, to clarify the qualitative
features we will ignore the imaginary parts of both dielectric
permittivities $\varepsilon _{0}$ and $\varepsilon _{1}(\omega )$. The
effect of the imaginary parts on the energy losses of the charge in the
problem at hand will be considered in Section \ref{sec:EnLoss} below. The
dispersion described by (\ref{eps1}) is the most popular model in
theoretical considerations of SPs. Putting $\eta =0$ in (\ref{eps1}), for
the upper frequency of SPs one gets $\omega <\omega _{p}$ and for the charge
velocity we have $\beta <1/\sqrt{\varepsilon _{0}}$. In terms of the
variable (\ref{u}) this constraint is reduced to $u<r_{c}\omega _{p}/v$.
From the features of the distribution of the roots $u_{n,s}$, described
above for general real $\varepsilon _{0}$ and $\varepsilon _{1}$, it follows
that for a given $\beta _{0}$ and $r_{c}\omega _{p}/c\gg 1$ for the roots
corresponding to $\omega /\omega _{p}$ one has $\omega /\omega _{p}\approx 1/%
\sqrt{1+\varepsilon _{0}/\gamma _{0}^{2}}$. In the opposite limit $%
r_{c}\omega _{p}/c\ll 1$ the roots with respect to $\omega /\omega _{p}$
tend to 1 for $n=0$ and to $1/\sqrt{1+\varepsilon _{0}}$ for $n\geq 1$.
Again, based on the general analysis presented above, we can see that for
fixed $r_{c}\omega _{p}/c$ and for large values of $n$ one has $\varepsilon
_{1}(\omega )/\varepsilon _{0}\approx -1$ or, in terms of the angular
frequency, $\omega /\omega _{p}\approx 1/\sqrt{1+\varepsilon _{0}}$. In
accordance of the interpretation given above, specified for the special case
at hand, the radiation modes $u=u_{n,s}$ are determined by the intersection
of the curves $\varepsilon _{1}=1-\left( r_{c}\omega _{p}/v\right)
^{2}/u^{2} $ and $f_{n}(u,\beta _{0})$. As seen from Figure \ref{fig2}, for
the corresponding example there is a unique solution and we can omit the
index $s $ for $u_{n,s}$ and the summation over $s$ in the expressions for
the fields given above.

\begin{figure}[tbph]
\begin{center}
\begin{tabular}{cc}
\epsfig{figure=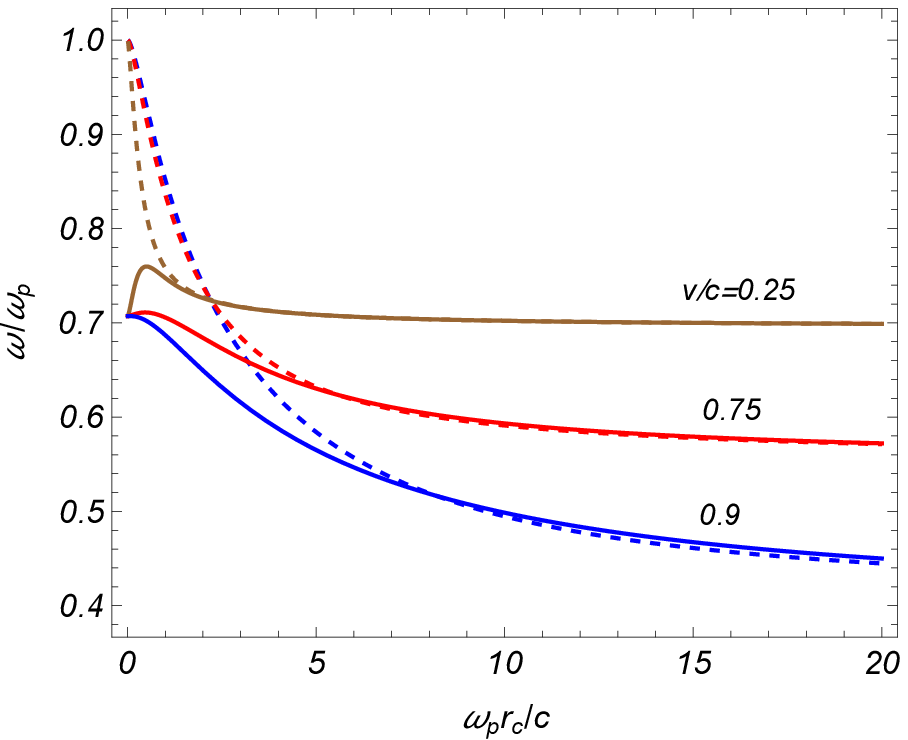,width=7.5cm,height=6.cm} & \quad %
\epsfig{figure=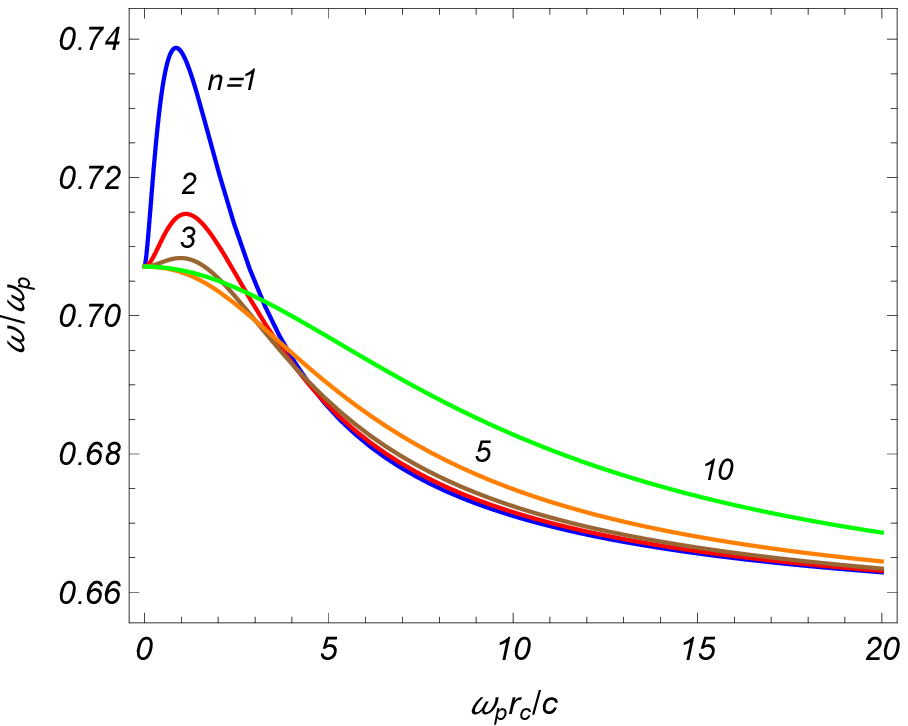,width=7.5cm,height=6.cm}%
\end{tabular}%
\end{center}
\caption{The frequencies of SP modes versus the combination $\protect\omega %
_{p}r_{c}/c$ in the case of the Drude model of dispersion for $\protect%
\varepsilon _{1}(\protect\omega )$ and for $\protect\varepsilon _{0}=1$. The
dashed and full curves on the left panel correspond to the modes $n=0$ and $%
n=1$, respectively. The numbers near the curves are the corresponding values
of $\protect\beta $. The graphs on the right panel are plotted for different
values of $n$ (the numbers near the curves) and for $\protect\beta =0.5$. }
\label{fig3}
\end{figure}

In Figure \ref{fig3} the roots of the eigenvalue equation for SPs with
respect to the ratio $\omega /\omega _{p}$ are depicted as functions of the
combination $\omega _{p}r_{c}/c$ for the dispersion law (\ref{eps1})
ignoring the absorption. For the region $r<r_{c}$ we have taken $\varepsilon
_{0}=1$. The left panel presents the curves for the modes $n=0$ (dashed
lines) and $n=1$ (full lines). The numbers near the curves are the values of 
$\beta $. The right panel displays the location of the roots for $\beta =0.5$
and for the modes $n=1,2,3,5,10$ (the numbers near the curves). The
numerical data confirm the asymptotic analysis given above.

\begin{figure}[tbph]
\begin{center}
\begin{tabular}{cc}
\epsfig{figure=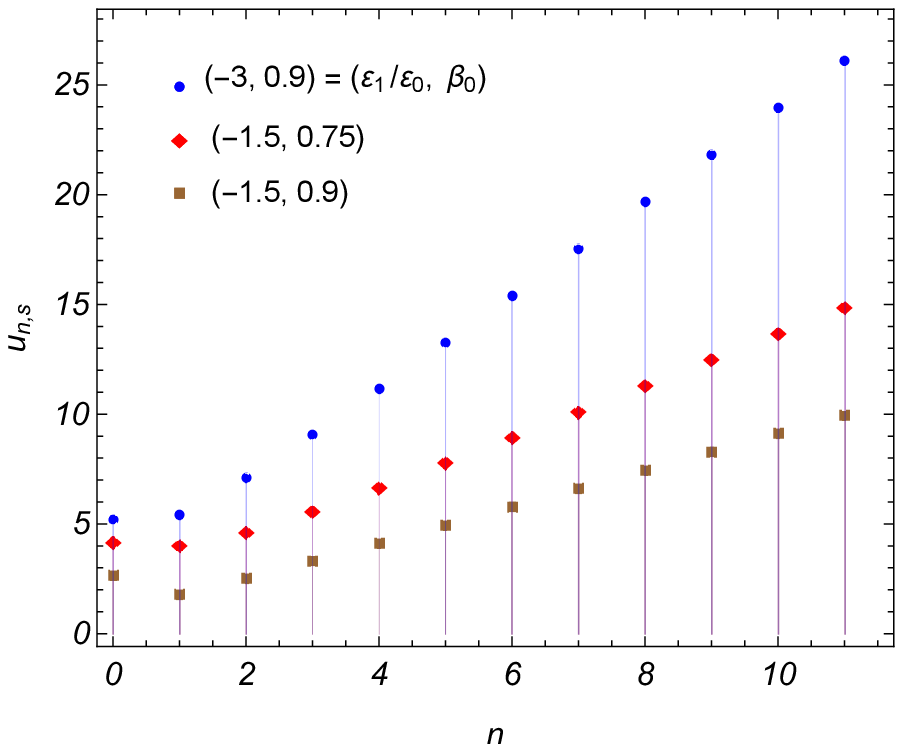,width=7.5cm,height=6.cm} & \quad %
\epsfig{figure=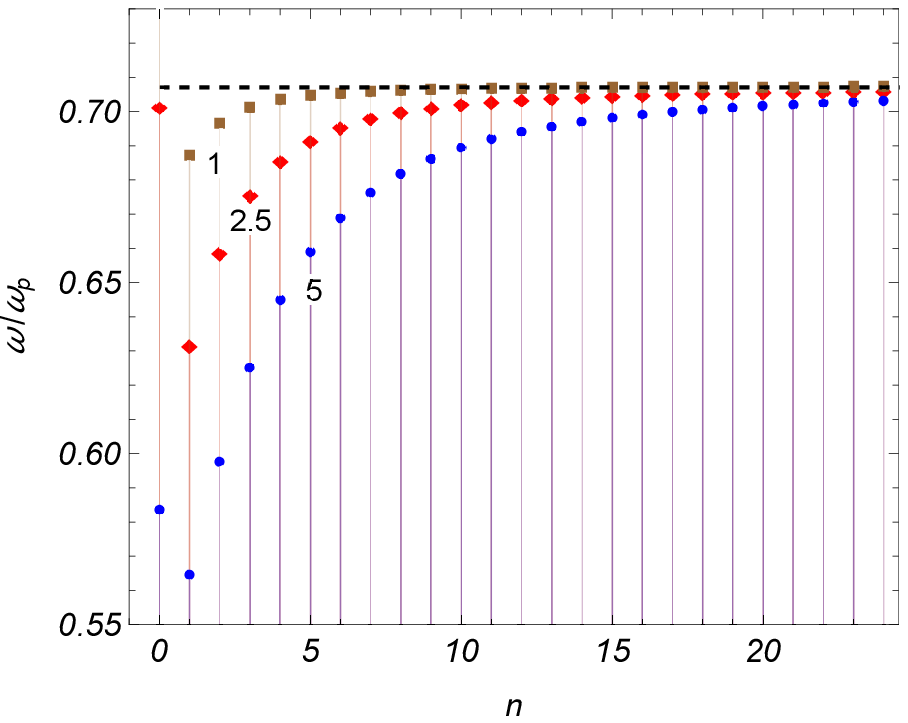,width=7.5cm,height=6.cm}%
\end{tabular}%
\end{center}
\caption{SP modes as functions of $n$ for the ratio $r_{0}/r_{c}=0.95$. The
left panel presents the roots with respect to $u=k_{z}r_{c}$ for given
values of the pair $(\protect\varepsilon _{1}/\protect\varepsilon _{0},%
\protect\beta _{0})$. On the right panel the modes with respect to the ratio 
$\protect\omega /\protect\omega _{p}$ are depicted in the Drude model (%
\protect\ref{eps1}) with $\protect\eta =0$. The numbers near the points are
the values of $r_{c}\protect\omega _{p}/c$.}
\label{fig4}
\end{figure}

As it has been discussed above, for large values of $n$ we can have two
qualitatively different cases for the behavior of the modes $u_{n,s}$. For
given $\varepsilon _{0}$ and $\varepsilon _{1}$, when $\varepsilon _{1}$ is
not too close to $-\varepsilon _{0}$, the roots linearly increase with
increasing $n$ (see (\ref{unsln})). This type of the behavior for $%
r_{0}/r_{c}=0.95$ is illustrated on the left panel of Figure \ref{fig4}. The
corresponding values for the pair $(\varepsilon _{1}/\varepsilon _{0},\beta
_{0})$ are presented on the figure. In the second case, the permittivity $%
\varepsilon _{1}$ is close to $-\varepsilon _{0}$ (see (\ref{alfln1})) and
for large $n$ one has $u_{n,s}\ll n$. This case is realized, for example, by
the dispersion law (\ref{eps1}) with $\eta =0$ and for the motion of the
charge in vacuum, $\varepsilon _{0}=1$. The results of the corresponding
numerical evaluations for $\omega /\omega _{p}$, with the values of the
parameters $r_{0}/r_{c}=0.95$, $\beta =0.9$, are presented on the right
panel of Figure \ref{fig4}. The numbers near the points correspond to the
values of $r_{c}\omega _{p}/c$. For large $n$ the ratio $\omega /\omega _{p}$
tends to $1/\sqrt{2}$ (dashed line, corresponding to $1/\sqrt{1+\varepsilon
_{0}}$ in the asymptotic analysis given above) and $\varepsilon _{1}(\omega
) $ tends to $-\varepsilon _{0}=-1$. On the right panel of Figure \ref{fig4}%
, for the $n=0$ mode in the case $r_{c}\omega _{p}/c=1$ one has $\omega
/\omega _{p}\approx 0.852$. With decreasing $\beta $ the distribution of the
modes near the line $1/\sqrt{2}$ becomes narrower. This feature can also be
seen from Figure \ref{fig3}.

\begin{figure}[tbph]
\begin{center}
\begin{tabular}{cc}
\epsfig{figure=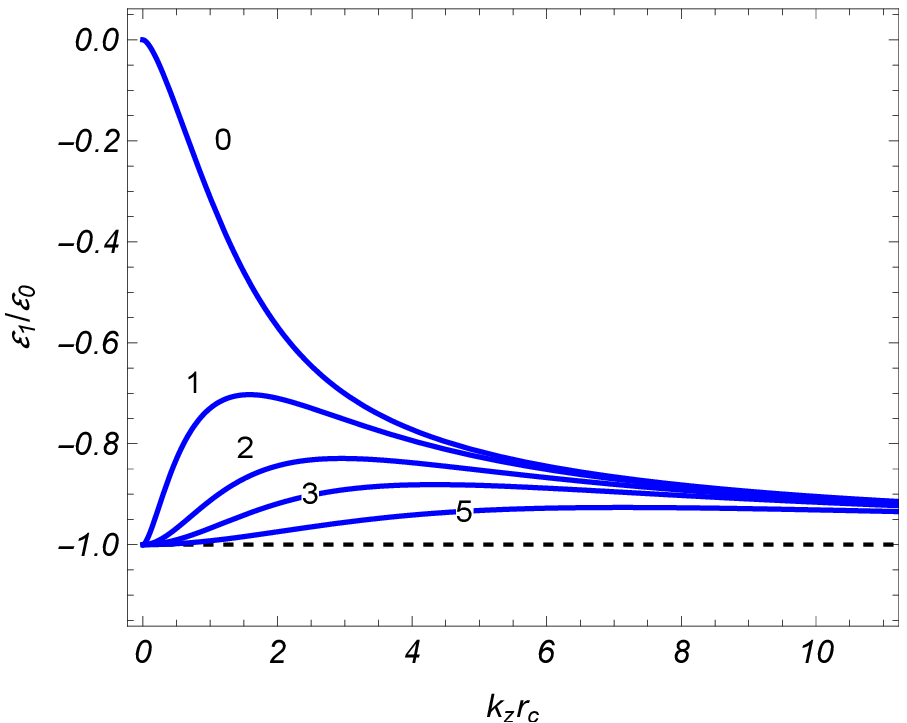,width=7.5cm,height=6.cm} & \quad %
\epsfig{figure=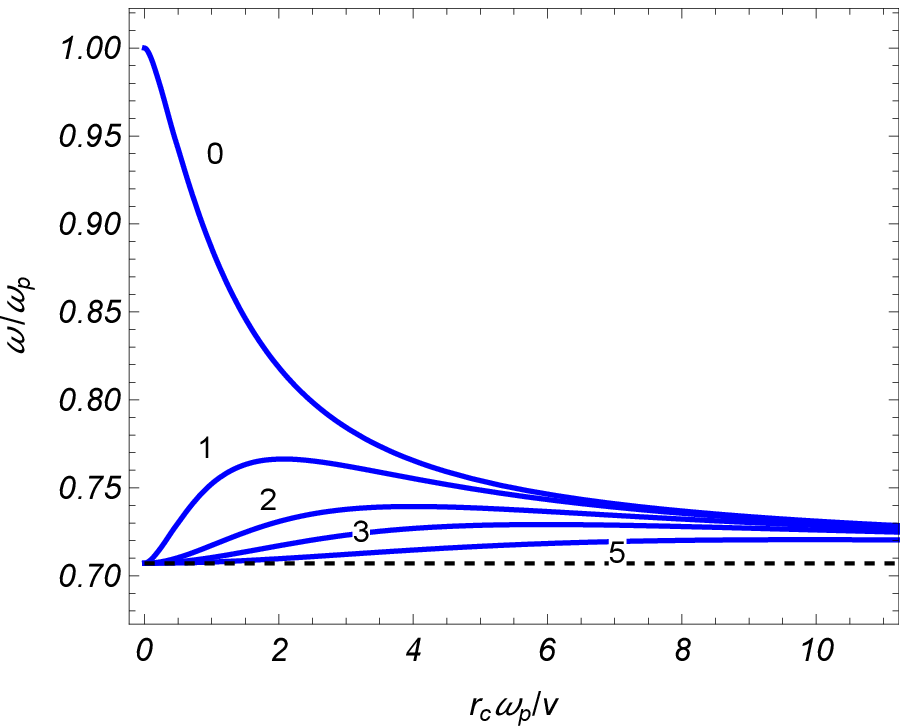,width=7.5cm,height=6.cm}%
\end{tabular}%
\end{center}
\caption{The left panel presents the function $\protect\varepsilon _{1}/%
\protect\varepsilon _{0}=f_{n}(k_{z}r_{c})$ for SPs in the non-relativistic
limit of the charge motion. The right panel describes the dependence of the
ratio $\protect\omega /\protect\omega _{p}$ on the quantity $r_{c}\protect%
\omega _{p}/v$ for the Drude model of dispersion in the same limit. The
numbers near the curves are the corresponding values for $n$. }
\label{fig5}
\end{figure}

In the non-relativistic limit, assuming that $\beta ^{2}|\varepsilon
_{j}|\ll 1$, to the leading order, we can replace $\gamma _{0}$ and $\gamma
_{1}$ in (\ref{SPmodes}) by 1. The corresponding expression is simplified by
using the Wronskian relation for the modified Bessel function and we get%
\begin{equation}
\alpha _{n}(u)\approx \frac{\varepsilon _{0}}{\varepsilon _{1}-\varepsilon
_{0}}-uI_{n}(u)K_{n}^{\prime }(u).  \label{alfnr}
\end{equation}%
Hence, in the non-relativistic limit the equation determining the eigenmodes
for the SPs reads%
\begin{equation}
uI_{n}(u)K_{n}^{\prime }(u)=\frac{\varepsilon _{0}}{\varepsilon
_{1}-\varepsilon _{0}}.  \label{Modesnr}
\end{equation}%
This equation can also be written in the form $uI_{n}^{\prime
}(u)K_{n}(u)=\varepsilon _{1}/(\varepsilon _{1}-\varepsilon _{0})$.
Combining these two forms we get another equivalent representation%
\begin{equation}
\frac{\varepsilon _{1}}{\varepsilon _{0}}=\frac{I_{n}^{\prime }(u)K_{n}(u)}{%
I_{n}(u)K_{n}^{\prime }(u)}.  \label{Modesnr2}
\end{equation}%
This form has been used, for example, in \cite{Aris01}. Similar to the
interpretation given above, we can consider (\ref{Modesnr}) as an equation
that determines the ratio $\varepsilon _{1}/\varepsilon _{0}$ as a function
of $u$ for a given $n$: $\varepsilon _{1}/\varepsilon _{0}=f_{n}(u)$. By
making use of the properties of the modified Bessel functions we can see
that $\lim_{u\rightarrow \infty }f_{n}(u)=-1$ for all $n$, $f_{0}(0)=0$, and 
$f_{n}(0)=-1$ for $n>0$. In addition we have $f_{n}(u)>f_{n+1}(u)$ and $%
-1<f_{n}(u)<0$ for $0<u<\infty $. The function $f_{0}(u)$ is monotonically
decreasing, whereas the functions $f_{n}(u)$ with $n>0$ have the maximum $%
f_{n}^{\mathrm{(m)}}<0$, $0\leq f_{n}(u)\leq f_{n}^{\mathrm{(m)}}$, which
decreases with increasing $n$. From this analysis we conclude that in the
non-relativistic limit SP modes are present in the range $-1\leq \varepsilon
_{1}/\varepsilon _{0}<0$ for $n=0$ and in the range $-1\leq \varepsilon
_{1}/\varepsilon _{0}\leq f_{n}^{\mathrm{(m)}}$ for $n>0$. The allowed
region for $\varepsilon _{1}/\varepsilon _{0}$ becomes narrower with
increasing $n$. All those features are seen from the left panel of Figure %
\ref{fig5} where we have displayed the function $\varepsilon
_{1}/\varepsilon _{0}=f_{n}(u)$ for different values of $n$ (the numbers
near the curves). The right panel in Figure \ref{fig5} presents the
dependence of the frequency of the SPs on the parameter $r_{c}\omega _{p}/v$
for the model of dispersion (\ref{eps1}) (with $\eta =0$) and for $%
\varepsilon _{0}=1$ in the non-relativistic limit. The numbers near the
curves are the values of $n$. In the limit $r_{c}\omega _{p}/v\rightarrow 0$
we have $\omega /\omega _{p}\rightarrow 0$ for $n=0$ and $\omega /\omega
_{p}\rightarrow 1/\sqrt{2}$ for $n>0$. In the opposite limit $r_{c}\omega
_{p}/v\rightarrow \infty $ we get $\omega /\omega _{p}\rightarrow 1/\sqrt{2}$%
. Comparing with the data presented in Figures \ref{fig2} and \ref{fig3} we
see that the relativistic effects may essentially enlarge the regions for $%
\varepsilon _{1}/\varepsilon _{0}$ in the general case and for $\omega
/\omega _{p}$ in the Drude model allowing the existence of the SP modes.

\subsection{Fields for a charge moving along the cylinder axis}

The expressions for the electromagnetic fields corresponding to the
generated SPs are essentially simplified in the special case of the charge
motion along the axis of the cylinder corresponding to $r_{0}=0$. The
dependence of the fields on $r_{0}$ enters in the expressions for the fields
in the form of the modified Bessel function $I_{n}(\gamma _{0}ur_{0}/r_{c})$
in (\ref{Qn}). From here we conclude that the only nonzero contribution
comes from the mode with $n=0$ and the fields do not depend on the angular
coordinate $\phi $. Of course, that is a direct consequence of the problem
symmetry in the special case at hand. For the function (\ref{alf2}) one gets%
\begin{equation}
\alpha _{0}(u)=\frac{\varepsilon _{0}\gamma _{1}I_{1}(\gamma
_{0}u)K_{0}(\gamma _{1}u)+\varepsilon _{1}\gamma _{0}I_{0}(\gamma
_{0}u)K_{1}(\gamma _{1}u)}{\left( \varepsilon _{0}-\varepsilon _{1}\right)
W_{1}^{I}},  \label{alf0}
\end{equation}%
and the equation determining the eigenmodes for SPs take the form%
\begin{equation}
\frac{I_{1}(\gamma _{0}u)K_{0}(\gamma _{1}u)}{I_{0}(\gamma
_{0}u)K_{1}(\gamma _{1}u)}=-\frac{\varepsilon _{1}\gamma _{0}}{\varepsilon
_{0}\gamma _{1}}.  \label{Modesn0}
\end{equation}%
In the non-relativistic limit the latter is reduced to (\ref{Modesnr2}) with 
$n=0$. Note that we have $0<\gamma _{0}<1<\gamma _{1}$. Under these
conditions the function in the left-hand side of (\ref{Modesn0}) is
monotonically increasing from 0 for $u=0$ and approaching 1 in the limit $%
u\rightarrow \infty $. From here it follows that the SP modes are present
under the condition $1/(\beta _{0}^{2}-1)<\varepsilon _{1}/\varepsilon
_{0}<0 $ and there is a single mode for given $\beta _{0}^{2}$ and $%
\varepsilon _{1}/\varepsilon _{0}$ obeying that condition.

From the general formulas, by using the equation (\ref{Modesn0}), for the
nonzero components of the potentials we get%
\begin{eqnarray}
\varphi ^{\mathrm{(P)}}(x) &=&\frac{q}{r_{c}}Q(u)R_{(0)}(u,r/r_{c})\sin
\left( u\xi /r_{c}\right) ,  \notag \\
A_{1}^{\mathrm{(P)}}(x) &=&-\frac{q}{r_{c}}\beta \varepsilon
Q(u)R_{(1)}(u,r/r_{c})\cos \left( u\xi /r_{c}\right) ,  \label{phin0}
\end{eqnarray}%
where $\varepsilon =\varepsilon _{0}$ for $r<r_{c}$, $\varepsilon
=\varepsilon _{1}$ for $r>r_{c}$, and 
\begin{equation}
Q(u)=\frac{2\varepsilon _{1}\left( \varepsilon _{0}-\varepsilon _{1}\right)
^{-2}\gamma _{0}}{uI_{1}(\gamma _{0}u)\bar{\alpha}_{0}(u)}.  \label{Qu}
\end{equation}%
The dependence on the radial coordinate is expressed in terms of the
functions%
\begin{equation}
R_{(0)}(u,w)=\left\{ 
\begin{array}{cc}
\frac{I_{0}(\gamma _{0}uw)}{I_{0}(\gamma _{0}u)}, & w<1 \\ 
\frac{K_{0}(\gamma _{1}uw)}{K_{0}(\gamma _{1}u)}, & w>1%
\end{array}%
\right. ,  \label{R0}
\end{equation}%
and%
\begin{equation}
R_{(1)}(u,w)=\left\{ 
\begin{array}{cc}
\frac{I_{1}(\gamma _{0}uw)}{\gamma _{0}I_{0}(\gamma _{0}u)}, & w<1 \\ 
-\frac{K_{1}(\gamma _{1}uw)}{\gamma _{1}K_{0}(\gamma _{1}u)}, & w>1%
\end{array}%
\right. .  \label{R1}
\end{equation}

For the nonzero components of the electric and magnetic fields one finds%
\begin{eqnarray}
E_{1}^{\mathrm{(P)}}(x) &=&\frac{H_{2}^{\mathrm{(P)}}(x)}{\beta \varepsilon }%
=-\frac{q}{r_{c}^{2}}uQ(u)R_{(1)}(u,r/r_{c})\sin \left( u\xi /r_{c}\right) ,
\notag \\
E_{3}^{\mathrm{(P)}}(x) &=&\frac{q}{r_{c}^{2}}uQ(u)R_{(0)}(u,r/r_{c})\cos
\left( u\xi /r_{c}\right) ,  \label{E3axial}
\end{eqnarray}%
where $u$ is a root of the equation (\ref{Modesn0}). For the special case of
axial motion the electric and magnetic fields are orthogonal and the
magnetic field is transversal.

\section{Energy fluxes for radiated surface polaritons}

\label{sec:Flux}

Having the expressions of the fields for radiated SPs, in this section we
evaluate the energy flux through the plane perpendicular to the cylinder
axis. The latter is determined by the Poynting vector and is given by 
\begin{equation}
I^{(f)}=\frac{c}{4\pi }\int_{0}^{2\pi }d\phi \int_{0}^{\infty }dr\,r\left[ 
\mathbf{E}^{\mathrm{(P)}}\times \mathbf{H}^{\mathrm{(P)}}\right] \cdot 
\mathbf{n}_{z},  \label{If}
\end{equation}%
with $\mathbf{n}_{z}$ being the unit vector along the $z$-axis. We will
evaluate the fluxes in the interior and exterior regions with the
integrations over $r\in \lbrack 0,r_{c}]$ and $r\in \lbrack r_{c},\infty )$,
respectively. Substituting the Fourier expansions (\ref{H3P}) and (\ref{ElP}%
), after integration over $\phi $ we find%
\begin{eqnarray}
I^{(f)}\left( \xi \right) &=&-\frac{c}{4}\int_{0}^{\infty
}dr\,r\sum_{n=0}^{\infty }\sum_{s,s^{\prime }}\sin \left( u_{n,s}\xi
/r_{c}\right) \sin \left( u_{n,s^{\prime }}\xi /r_{c}\right)  \notag \\
&&\times \left[ \left( 1+\delta _{0n}\right) E_{1,n}^{\mathrm{(P)}}\left(
u_{n,s}\right) H_{2,n}^{\mathrm{(P)}}\left( u_{n,s^{\prime }}\right) +\left(
1-\delta _{0n}\right) E_{2,n}^{\mathrm{(P)}}\left( u_{n,s}\right) H_{1,n}^{%
\mathrm{(P)}}\left( u_{n,s^{\prime }}\right) \right] ,  \label{Iksi}
\end{eqnarray}%
where the Fourier components are given by (\ref{H3Pn}) and (\ref{ElPn}).

The energy flux through the plane $z=\mathrm{const}$ during the time
interval $z/v\leq t\leq t_{0}$ is given by $\mathcal{E}_{[z/v,t_{0}]}^{(f)}=%
\int_{z/v}^{t_{0}}dt\,I^{(f)}\left( \xi \right) =\int_{0}^{vt_{0}-z}d\xi
\,I^{(f)}\left( \xi \right) /v$. The difference of the corresponding fluxes
for the planes $z=z_{1}$ and $z=z_{2}>z_{1}$ is expressed as 
\begin{equation}
\mathcal{E}^{(f)}(t_{0},z_{1},z_{2})=\mathcal{E}_{[t_{1},t_{0}]}^{(f)}-%
\mathcal{E}_{[t_{2},t_{0}]}^{(f)}=\frac{1}{v}%
\int_{vt_{0}-z_{2}}^{vt_{0}-z_{1}}d\xi \,I^{(f)}\left( \xi \right) ,
\label{Ecalf}
\end{equation}%
where $t_{j}=z_{j}/v$, $j=1,2$. This difference characterizes the energy
radiated by the charge during the time interval $t\in \lbrack t_{1},t_{2}]$.
In the limit $t_{0}\rightarrow \infty $ we use the result%
\begin{equation}
\lim_{t_{0}\rightarrow \infty }\int_{vt_{0}-z_{2}}^{vt_{0}-z_{1}}d\xi \,\sin
\left( u\xi /r_{c}\right) \sin \left( u^{\prime }\xi /r_{c}\right) =\frac{%
t_{2}-t_{1}}{2v}\delta _{u^{\prime }u}.  \label{rel1}
\end{equation}%
By using (\ref{Iksi}), for the mean energy flux (averaged in the way
described above) per unit time through the plane $z=\mathrm{const}$, given
by $I^{(f)}=\lim_{t_{0}\rightarrow \infty }\mathcal{E}%
^{(f)}(t_{0},z_{1},z_{2})/(t_{2}-t_{1})$, we get%
\begin{equation}
I^{(f)}=-\frac{c}{8}\int_{0}^{\infty }dr\,r\sum_{n=0}^{\infty }\sum_{s}\left[
\left( 1+\delta _{0n}\right) E_{1,n}^{\mathrm{(P)}}\left( u\right) H_{2,n}^{%
\mathrm{(P)}}\left( u\right) +\left( 1-\delta _{0n}\right) E_{2,n}^{\mathrm{%
(P)}}\left( u\right) H_{1,n}^{\mathrm{(P)}}\left( u\right) \right]
_{u=u_{n,s}}.  \label{Ifmean}
\end{equation}

We write the total energy flux in the form%
\begin{equation*}
I^{(f)}=\sum_{n=0}^{\infty }\sum_{s}\left( I_{\mathrm{i},n,s}^{(f)}+I_{%
\mathrm{e},n,s}^{(f)}\right) ,
\end{equation*}%
where $I_{\mathrm{i},n,s}^{(f)}$ and $I_{\mathrm{e},n,s}^{(f)}$ are the
energy fluxes on a given mode $u=u_{n,s}$ in the regions $r<r_{c}$
(interior) and $r>r_{c}$ (exterior). Those separate contributions are
obtained from (\ref{Ifmean}) by using the expressions for the Fourier
components of the electric and magnetic fields given in the previous
section. The corresponding radial integrals are reduced to \cite{Prud2} 
\begin{eqnarray}
\int_{0}^{r_{c}}dr\,rI_{n+p}^{2}(\gamma _{0}ur/r_{c}) &=&\frac{r_{c}^{2}}{2}%
\left[ I_{n+p}^{2}(\gamma _{0}u)-I_{n+2p}(\gamma _{0}u)I_{n}(\gamma _{0}u)%
\right] ,  \notag \\
\int_{r_{c}}^{\infty }dr\,rK_{n+p}^{2}(\gamma _{1}ur/r_{c}) &=&\frac{%
r_{c}^{2}}{2}\left[ K_{n+2p}(\gamma _{1}u)K_{n}(\gamma
_{1}u)-K_{n+p}^{2}(\gamma _{1}u)\right] .  \label{rint}
\end{eqnarray}%
For the energy flux in the region $r<r_{c}$ we find%
\begin{eqnarray}
I_{\mathrm{i},n,s}^{(f)} &=&\delta _{n}\frac{q^{2}v}{4r_{c}^{2}}\frac{%
Q_{n}^{2}(u)}{\varepsilon _{0}}\sum_{p,p^{\prime }=\pm 1}\left( 1+pp^{\prime
}\beta ^{2}\varepsilon _{0}\right) \frac{K_{n+p^{\prime }}(\gamma _{1}u)}{%
W_{n+p^{\prime }}^{I}}  \notag \\
&&\times \frac{K_{n+p}(\gamma _{1}u)}{W_{n+p}^{I}}\left[ I_{n+p}^{2}(\gamma
_{0}u)-I_{n+2p}(\gamma _{0}u)I_{n}(\gamma _{0}u)\right] |_{u=u_{n,s}}.
\label{Ifi}
\end{eqnarray}%
Note that by using the first relation in (\ref{RelK}) one has%
\begin{equation}
\sum_{p^{\prime }=\pm 1}\left( 1+pp^{\prime }\beta ^{2}\varepsilon
_{0}\right) \frac{K_{n+p^{\prime }}(\gamma _{1}u)}{W_{n+p^{\prime }}^{I}}%
=2p\varepsilon _{0}\left[ \beta ^{2}\frac{K_{n+1}(\gamma _{1}u)}{W_{n+1}^{I}}%
-\frac{p-\beta _{0}^{2}}{\left( \varepsilon _{0}-\varepsilon _{1}\right)
\gamma _{0}I_{n}(\gamma _{0}u)}\right] .  \label{RelK2}
\end{equation}%
By taking into account that $W_{n+p}^{I}<0$, from (\ref{Ifi}) we see that
the energy flux is always positive.

In a similar way, for the energy flux of SPs in the exterior region, $%
r>r_{c} $, one gets 
\begin{eqnarray}
I_{\mathrm{e},n,s}^{(f)} &=&\delta _{n}\frac{q^{2}v}{4r_{c}^{2}}\frac{%
Q_{n}^{2}(u)}{\varepsilon _{1}}\sum_{p,p^{\prime }=\pm 1}\left( 1+pp^{\prime
}\beta ^{2}\varepsilon _{1}\right) \frac{I_{n+p^{\prime }}(\gamma _{0}u)}{%
W_{n+p^{\prime }}^{I}}  \notag \\
&&\times \frac{I_{n+p}(\gamma _{0}u)}{W_{n+p}^{I}}\left[ K_{n+2p}(\gamma
_{1}u)K_{n}(\gamma _{1}u)-K_{n+p}^{2}(\gamma _{1}u)\right] |_{u=u_{n,s}}.
\label{Ife}
\end{eqnarray}%
From the second relation in (\ref{RelK}) we have%
\begin{equation}
\sum_{p^{\prime }=\pm 1}\left( 1+pp^{\prime }\beta ^{2}\varepsilon
_{1}\right) \frac{I_{n+p^{\prime }}(\gamma _{0}u)}{W_{n+p^{\prime }}^{I}}%
=2p\varepsilon _{1}\left[ \beta ^{2}\frac{I_{n+1}(\gamma _{0}u)}{W_{n+1}^{I}}%
+\frac{p-\beta ^{2}\varepsilon _{1}}{\left( \varepsilon _{0}-\varepsilon
_{1}\right) \gamma _{1}K_{n}(\gamma _{1}u)}\right] .  \label{RelI2}
\end{equation}%
For the average total energy flux on a given mode $k_{n,s}=u_{n,s}/r_{c}$ we
have%
\begin{equation}
I_{\mathrm{t},n,s}^{(f)}=I_{\mathrm{i},n,s}^{(f)}+I_{\mathrm{e},n,s}^{(f)}.
\label{Ift}
\end{equation}%
By taking into account that for given values of $\beta _{0}$ and $%
\varepsilon _{1}/\varepsilon _{0}$ the roots do not depend on $r_{c}$ and $%
r_{0}$, we see that the dependence of the energy fluxes on those parameters
appears in the form $I_{n}^{2}(\gamma _{0}ur_{0}/r_{c})/r_{c}^{2}$. In
particular, for a fixed ratio $r_{0}/r_{c}$ the fluxes decay as $1/r_{c}^{2}$
with increasing radius of the cylinder. In the non-relativistic limit, to
the leading order, the modes $u_{n,s}$ are roots of the equation (\ref%
{Modesnr2}) and for given $\varepsilon _{1}/\varepsilon _{0}$ do not depend
on the charge velocity. In the same order, one has $\gamma _{j}\approx 1$, $%
j=0,1$, and from (\ref{Ifi}) and (\ref{Ife}) we conclude that the energy
fluxes behave as $I_{\mathrm{j},n,s}^{(f)}\propto \beta $, $\mathrm{j}=%
\mathrm{i},\mathrm{e}$, for $\beta \ll 1$.

Let us consider the asymptotic behavior of the energy fluxes for large
values $u=u_{n,s}$. As it has been discussed above, this asymptotic is
realized in the range of dielectric permittivities where 
\begin{equation}
\varepsilon _{1}/\varepsilon _{0}\approx -\gamma _{0}^{-2}.  \label{epsrat}
\end{equation}%
For the asymptotic of the function $\bar{\alpha}_{n}(u)$ at the points $%
u=u_{n,s}$ we get 
\begin{equation}
\bar{\alpha}_{n}(u)\approx \frac{\varepsilon _{1}/\varepsilon
_{0}+\varepsilon _{0}/\varepsilon _{1}+1+2\beta _{0}^{2}u/\gamma _{0}}{%
2\gamma _{0}\left( 1-\varepsilon _{1}/\varepsilon _{0}\right) u^{2}}.
\label{alfndas}
\end{equation}%
Note that in the numerator we have kept additional terms to include in the
asymptotic analysis the region $\beta _{0}^{2}u\lesssim 1$. By using the
asymptotic expressions of the modified Bessel functions for large argument 
\cite{Abra72} it can be seen that, to the leading order,%
\begin{equation}
I_{\mathrm{i},n,s}^{(f)}\approx \frac{4\delta _{n}q^{2}v}{%
r_{0}r_{c}\varepsilon _{0}}\frac{\left( 1-\varepsilon _{0}/\varepsilon
_{1}\right) ^{-2}u_{n,s}^{3}\gamma _{0}e^{-2\left( 1-r_{0}/r_{c}\right)
\gamma _{0}u_{n,s}}}{\left( \varepsilon _{1}/\varepsilon _{0}+\varepsilon
_{0}/\varepsilon _{1}+1+2\beta _{0}^{2}u_{n,s}/\gamma _{0}\right) ^{2}}.
\label{Ifias}
\end{equation}%
In a similar way, for the energy flux in the exterior medium one gets%
\begin{equation}
I_{\mathrm{e},n,s}^{(f)}\approx -\frac{4\delta _{n}q^{2}v}{%
r_{0}r_{c}\varepsilon _{0}}\frac{\left( 1-\varepsilon _{1}/\varepsilon
_{0}\right) ^{-2}u_{n,s}^{3}\gamma _{0}e^{-2\left( 1-r_{0}/r_{c}\right)
\gamma _{0}u_{n,s}}}{\left( \varepsilon _{1}/\varepsilon _{0}+\varepsilon
_{0}/\varepsilon _{1}+1+2\beta _{0}^{2}u_{n,s}/\gamma _{0}\right) ^{2}}.
\label{Ifeas}
\end{equation}%
For the total energy flux this gives 
\begin{equation}
I_{\mathrm{t},n,s}^{(f)}\approx \frac{4\delta _{n}q^{2}v}{%
r_{0}r_{c}\varepsilon _{0}}\frac{u_{n,s}^{3}\gamma _{0}e^{-2\left(
1-r_{0}/r_{c}\right) \gamma _{0}u_{n,s}}}{\left( \varepsilon
_{1}/\varepsilon _{0}+\varepsilon _{0}/\varepsilon _{1}+1+2\beta
_{0}^{2}u_{n,s}/\gamma _{0}\right) ^{2}}\frac{\varepsilon _{1}+\varepsilon
_{0}}{\varepsilon _{1}-\varepsilon _{0}}.  \label{Ifnas}
\end{equation}%
The energy flux is positive/negative in the medium with positive/negative
dielectric permittivity. From (\ref{epsrat}) it follows that $\varepsilon
_{1}+\varepsilon _{0}<0$ and the total energy flux is positive. The exponent
in the asymptotic expressions (\ref{Ifias}) and (\ref{Ifeas}) is written as $%
2\left( 1-r_{0}/r_{c}\right) \gamma _{0}u_{n,s}=4\pi \left(
r_{c}-r_{0}\right) \gamma _{0}/\lambda _{n,s}$ and we see that the energy
fluxes for a given radiation wavelength are exponentially suppressed if the
distance of the charge trajectory from the cylinder surface is much larger
than the wavelength. The suppression factor decreases with increasing
velocity of the charge. We could expect the exponential suppression of
energy fluxes for large values of $u_{n,s}$. The SP for a given frequency is
generated by the corresponding spectral component of the charge proper
field. For a charge moving with constant velocity and at distances from the
trajectory larger than the wavelength the spectral component of the field is
exponentially small. The exponential factor in the asymptotic expressions (%
\ref{Ifias}) and (\ref{Ifeas}) is directly related the the corresponding
suppression factor in the proper field of the charge. Note that in obtaining
the asymptotics for large $u$ we have used approximate expressions for the
modified Bessel functions which are valid for $n\ll \gamma _{j}u$.

In order to see the features of the energy fluxes for large values of the
azimuthal quantum number $n$ we use the corresponding uniform asymptotic
expansions of the modified Bessel functions \cite{Abra72}. Those expansions
contain the exponential factor $e^{-n\eta (x/n)}$ for the function $K_{n}(x)$
and the factor $e^{n\eta (x/n)}$ for $I_{n}(x)$, where 
\begin{equation}
\eta (x)=\sqrt{1+x^{2}}+\ln \frac{x}{1+\sqrt{1+x^{2}}}.  \label{etax}
\end{equation}%
For the related exponential factors in the expressions of the energy fluxes
we get%
\begin{equation}
I_{\mathrm{j},n,s}^{(f)}\propto e^{2n\left[ \eta (\gamma
_{0}ur_{0}/nr_{c})-\eta (\gamma _{0}u/n)\right] },\;\mathrm{j}=\mathrm{i},%
\mathrm{e},  \label{Ijln}
\end{equation}%
with $u=u_{n,s}$. For $\gamma _{j}u\gg n$ the exponent in (\ref{Ijln}) is
reduced to the one in (\ref{Ifias}) and (\ref{Ifeas}). Note that $\eta
^{\prime }(x)=\sqrt{1+x^{2}}/x$ and the function $\eta (x)$ is monotonically
increasing for $x>0$. As a consequence, the exponent in (\ref{Ijln}) is
negative. As it has been already discussed in Section \ref{sec:AHE}, for
large $n$ two qualitatively different possibilities are realized. If for the
corresponding frequencies the permittivity $\varepsilon _{1}$ is
sufficiently close to $-\varepsilon _{0}$ (see (\ref{alfln1})), we have $%
u_{n,s}\ll n$ and the arguments of the functions $\eta (x)$ in (\ref{Ijln})
are small. In this regime, to the leading order one gets $I_{\mathrm{j}%
,n,s}^{(f)}\propto (r_{0}/r_{c})^{2n}$. In the second case, $\varepsilon
_{1} $ and $-\varepsilon _{0}$ are not too close and the modes $u_{n,s}$ are
approximated by (\ref{unsln}). The corresponding exponential factors in the
expressions for the energy fluxes are obtained from (\ref{Ijln}) by the
replacement $u/n\rightarrow 1/\sqrt{\beta _{0}^{2}\varepsilon
_{1}/(\varepsilon _{0}+\varepsilon _{1})-1}$.

In the special case of the axial motion the SPs are radiated only on the
mode $n=0$. For given $\beta _{0}$ and $\varepsilon _{1}/\varepsilon _{0}$
there is a single mode $u=u_{0}$ and it is the root of the equation (\ref%
{Modesn0}). By using that equation, the expressions for the interior and
exterior fluxes are simplified as%
\begin{equation}
I_{\mathrm{i},0}^{(f)}=\frac{q^{2}v}{2r_{c}^{2}\varepsilon _{0}}\frac{%
\varepsilon _{0}^{2}}{\varepsilon _{1}^{2}}\frac{\left[ \frac{I_{1}(\gamma
_{0}u)}{I_{0}(\gamma _{0}u)}+\frac{1}{\gamma _{0}u}\right] ^{2}-\frac{1}{%
\gamma _{0}^{2}u^{2}}-1}{\left( 1-\varepsilon _{0}/\varepsilon _{1}\right)
^{4}I_{1}^{2}(\gamma _{0}u)\bar{\alpha}_{0}^{2}(u)},  \label{Ii0}
\end{equation}%
and 
\begin{equation}
I_{\mathrm{e},0}^{(f)}=-\frac{q^{2}v}{2r_{c}^{2}\varepsilon _{1}}\frac{%
\gamma _{0}^{2}}{\gamma _{1}^{2}}\frac{\left[ \frac{K_{1}(\gamma _{1}u)}{%
K_{0}(\gamma _{1}u)}-\frac{1}{\gamma _{1}u}\right] ^{2}-\frac{1}{\gamma
_{1}^{2}u^{2}}-1}{\left( 1-\varepsilon _{0}/\varepsilon _{1}\right)
^{4}I_{1}^{2}(\gamma _{0}u)\bar{\alpha}_{0}^{2}(u)},  \label{Ie0}
\end{equation}%
with $u=u_{0}$. We can see that $I_{\mathrm{e},0}^{(f)}<0<I_{\mathrm{i}%
,0}^{(f)}$. By taking into account that the root $u_{0}$ does not depend on $%
r_{c}$, we conclude that for fixed values of the other parameters the energy
fluxes, as functions of the cylinder radius, behave like $1/r_{c}^{2}$.

\begin{figure}[tbph]
\begin{center}
\begin{tabular}{cc}
\epsfig{figure=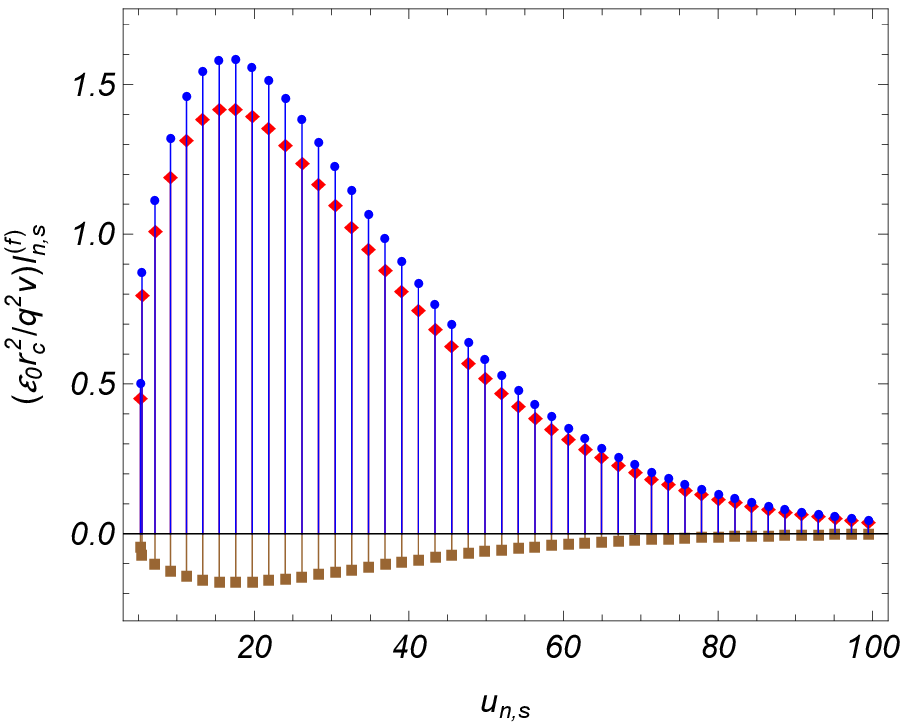,width=7.5cm,height=6.cm} & \quad %
\epsfig{figure=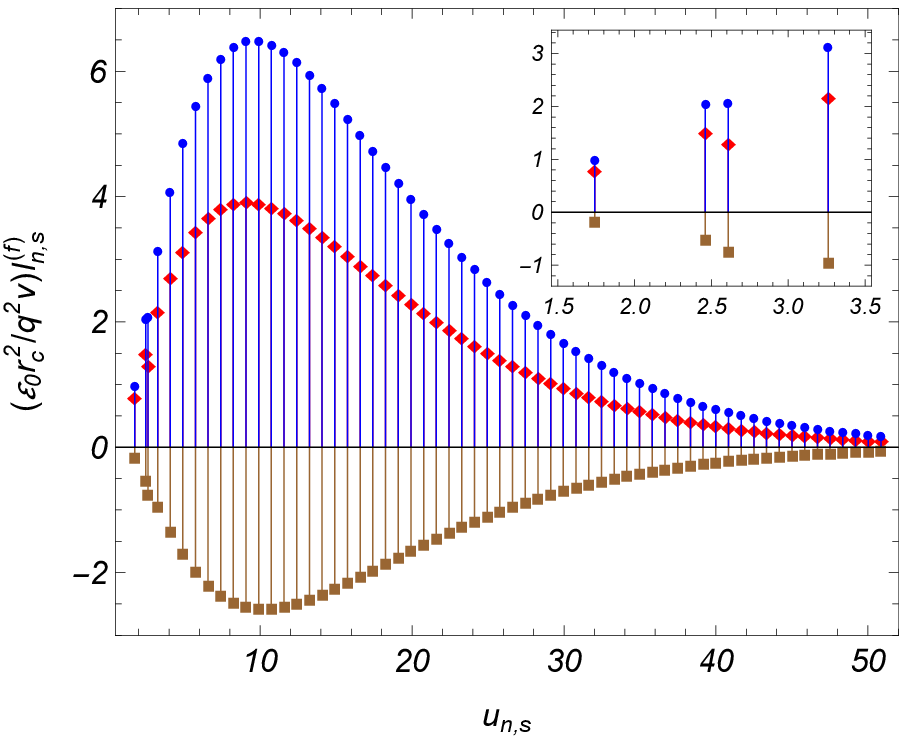,width=7.5cm,height=6.cm}%
\end{tabular}%
\end{center}
\caption{Energy fluxes of the radiated SPs inside (circles) and outside
(squares) the cylinder versus $u_{n,s}$. The points marked by diamonds
correspond to the total energy flux. The graphs are plotted for $\protect%
\beta _{0}=0.9$, $r_{0}/r_{c}=0.95$, and for two values of the ratio of
dielectric permittivities: $\protect\varepsilon _{1}/\protect\varepsilon %
_{0}=-3$ (left panel) and $\protect\varepsilon _{1}/\protect\varepsilon %
_{0}=-1.5$ (right panel).}
\label{fig6}
\end{figure}

We recall that for a given $n$ the roots $u_{n,s}$ are determined by $\beta
_{0}$ and $\varepsilon _{1}/\varepsilon _{0}$: $u_{n,s}=u_{n,s}(\beta
_{0},\varepsilon _{1}/\varepsilon _{0})$. From (\ref{Ifi}) and (\ref{Ife})
it is seen that the dimensionless combination $\varepsilon _{0}r_{c}^{2}I_{%
\mathrm{j},n,s}^{(f)}/(q^{2}v)$, with j=i,e,t, is completely determined by
the values $\beta _{0}$, $\varepsilon _{1}/\varepsilon _{0}$, and $%
r_{0}/r_{c}$. That combination corresponds to the energy flux radiated by
the charge from the part of trajectory of the length $r_{c}$, measured in
units of $q^{2}/\varepsilon _{0}r_{c}$. In Figure \ref{fig6} we display the
energy fluxes for radiated SPs in the exterior and interior regions as
functions of $u_{n,s}=2\pi r_{c}/\lambda _{n,s}$, with $\lambda _{n,s}$
being the wavelength. In the numerical evaluation we have taken $\beta
_{0}=0.9$ and $r_{0}/r_{c}=0.95$. The left and right panels correspond to $%
\varepsilon _{0}/\varepsilon _{1}=-3$ and $\varepsilon _{0}/\varepsilon
_{1}=-1.5$, respectively. The plot markers circles, squares and diamonds
correspond to the interior, $I_{\mathrm{i},n,s}^{(f)}$, exterior, $I_{%
\mathrm{e},n,s}^{(f)}$, and total, $I_{n,s}^{(f)}$, energy fluxes,
respectively. The modes $u_{n,s}$ on the horizontal axis of the left panel
correspond to the modes with $0\leq n\leq 45$, whereas on the right panel $%
0\leq n\leq 60$. In general, the roots $u_{n,s}$ are not monotonic functions
of $n$, though that is the case for large $n$. For example, in the case $%
\varepsilon _{0}/\varepsilon _{1}=-1.5$ one has $u_{0,s}\approx 2.61$, $%
u_{1,s}\approx 1.74$, and $u_{2,s}\approx 2.46$. As seen, the energy flux is
directed along the charge motion in the interior region (with positive
dielectric permittivity) and along the negative direction of $z$-axis in the
exterior region (with negative dielectric permittivity). The total energy
flux is dominated by the contribution of the interior region.

\begin{figure}[tbph]
\begin{center}
\begin{tabular}{cc}
\epsfig{figure=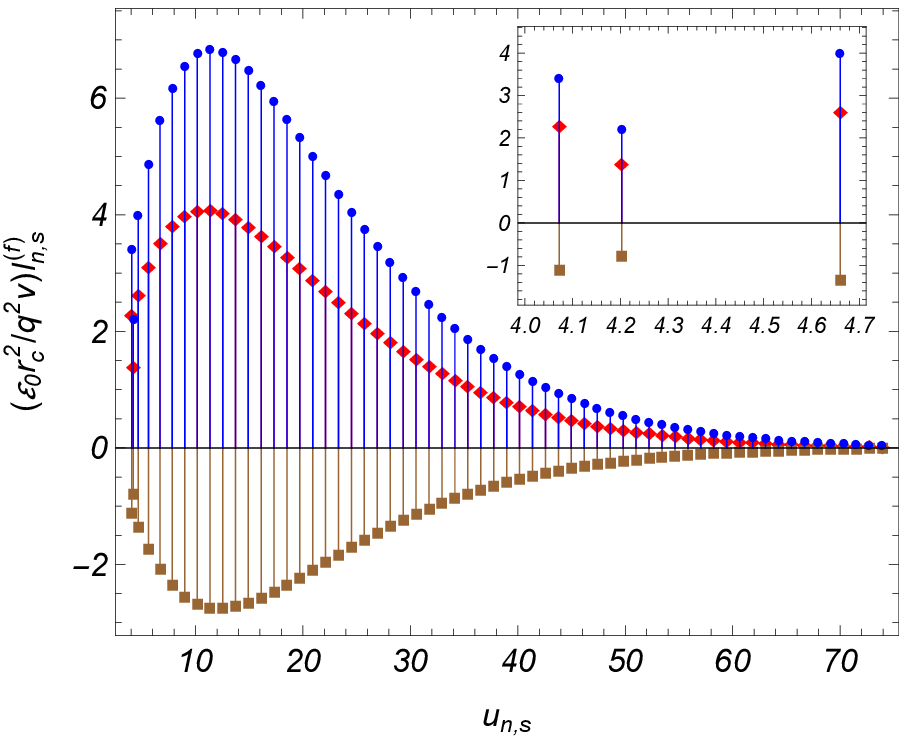,width=7.5cm,height=6.cm} & \quad %
\epsfig{figure=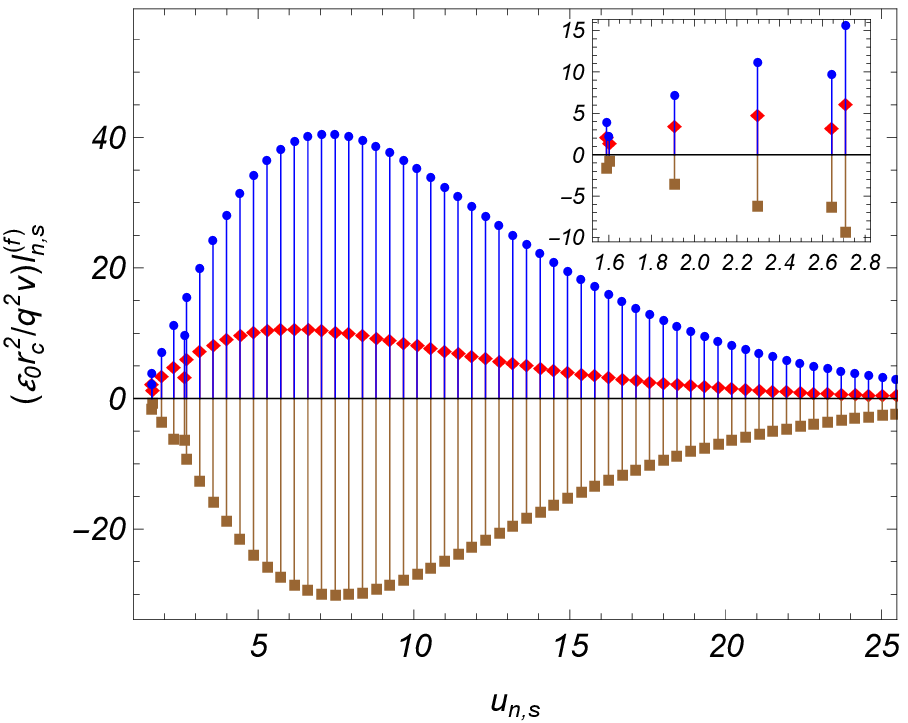,width=7.5cm,height=6.cm}%
\end{tabular}%
\end{center}
\caption{The same as in Figure \protect\ref{fig6} for $\protect\beta %
_{0}=0.75$. The left and right panels correspond to $\protect\varepsilon %
_{1}/\protect\varepsilon _{0}=-1.5$ and $\protect\varepsilon _{1}/\protect%
\varepsilon _{0}=-1.1$, respectively.}
\label{fig7}
\end{figure}

Figure \ref{fig7} presents the dependence of the energy fluxes on $u_{n,s}$
for $\beta _{0}=0.75$, $r_{0}/r_{c}=0.95$. The left panel is plotted for $%
\varepsilon _{1}/\varepsilon _{0}=-1.5$ and for the right panel $\varepsilon
_{1}/\varepsilon _{0}=-1.1$. As it has been already mentioned above, the
energy fluxes are exponentially suppressed for large values of $u_{n,s}$
corresponding to small wavelengths. The exponent of the suppression factor
is expressed as $2\left( 1-r_{0}/r_{c}\right) \gamma _{0}u_{n,s}$ and the
characteristic value of $u_{n,s}$, given by $1/[2\left( 1-r_{0}/r_{c}\right)
\gamma _{0}]$, is equal to $\approx 23$ and $\approx 15$ for the parameters
corresponding to Figures \ref{fig6} and \ref{fig7}. These numbers are in
agreement of the numerical data in figures.

\begin{figure}[tbph]
\begin{center}
\begin{tabular}{cc}
\epsfig{figure=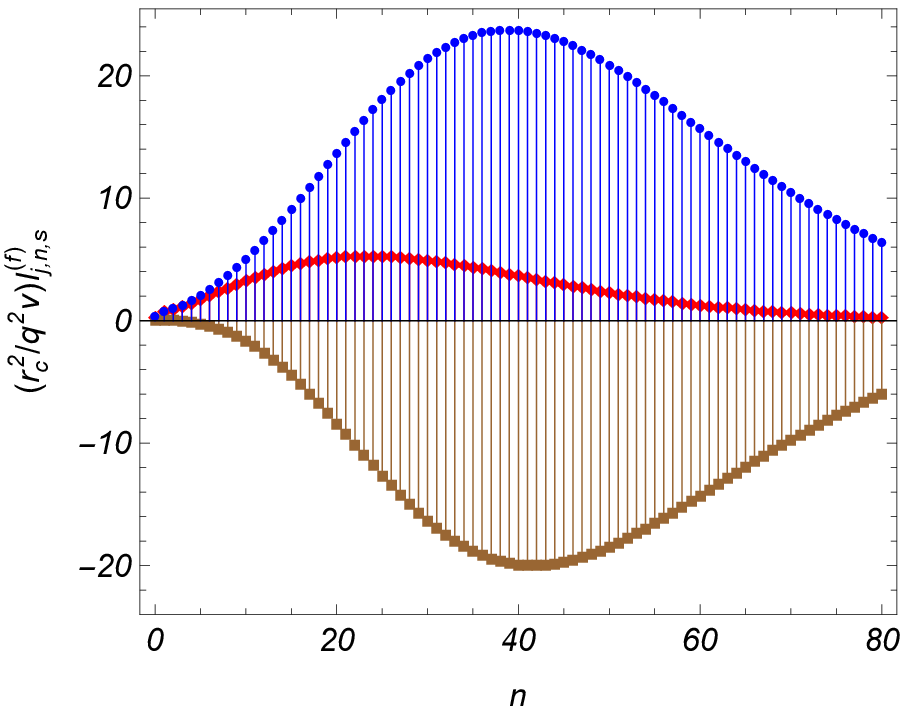,width=7.5cm,height=6.cm} & \quad %
\epsfig{figure=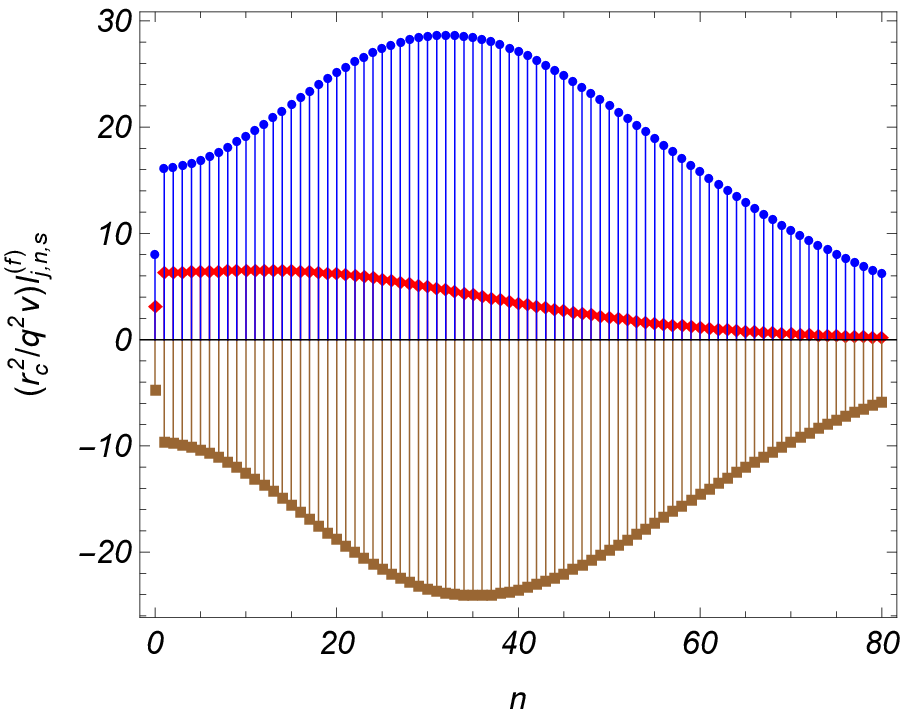,width=7.5cm,height=6.cm}%
\end{tabular}%
\end{center}
\caption{Energy fluxes for the SPs as functions of $n$ for the dielectric
function (\protect\ref{eps1}) and for $\protect\varepsilon _{0}=1$. The
points marked by circles, squares and diamonds correspond to the interior,
exterior and total energy fluxes. For the parameters we have taken $%
r_{0}/r_{c}=0.95$, $r_{c}\protect\omega _{p}/c=15$, and $\protect\beta %
_{0}=0.9,0.5$ for the left and right panels, respectively.}
\label{fig8}
\end{figure}

In Figure \ref{fig8} we have presented the energy fluxes for SPs versus $n$, 
$0\leq n\leq 80$, in the case when the dispersion of the dielectric function
for the medium in the region $r>r_{c}$ is described by (\ref{eps1}) and for
the interior region we have taken $\varepsilon _{0}=1$. The fluxes are
evaluated for $r_{c}\omega _{p}/c=15$ and $r_{0}/r_{c}=0.95$. The left and
right panels are plotted for $v/c=0.9$ and $v/c=0.5$, respectively. For the
example on the left panel the frequencies $\omega =\omega _{n}$
monotonically increase with increasing $n$ and quickly converges to the
limit $1/\sqrt{2}$ for large $n$. One has $\omega _{n}/\omega _{p}\approx
0.4611$ for $n=0$ and $\omega _{n}/\omega _{p}\approx 0.7041$ for $n=80$. We
have similar behavior in the example of the right panel with $\omega
_{n}/\omega _{p}\approx 0.6655$ for $n=0$ and $\omega _{n}/\omega
_{p}\approx 0.7043$ for $n=80$. In both cases the function $\varepsilon
_{1}(\omega )$ tends to $-1$. As it has been explained above, for the model
of dispersion at hand and for large $n$ the eigenfrequencies of the radiated
SPs are localized in the narrow range near the frequency $\omega _{p}/\sqrt{2%
}$. The data presented in Figure \ref{fig8} show that the main part of the
energy is radiated in that frequency range. 

\begin{figure}[tbph]
\begin{center}
\begin{tabular}{cc}
\epsfig{figure=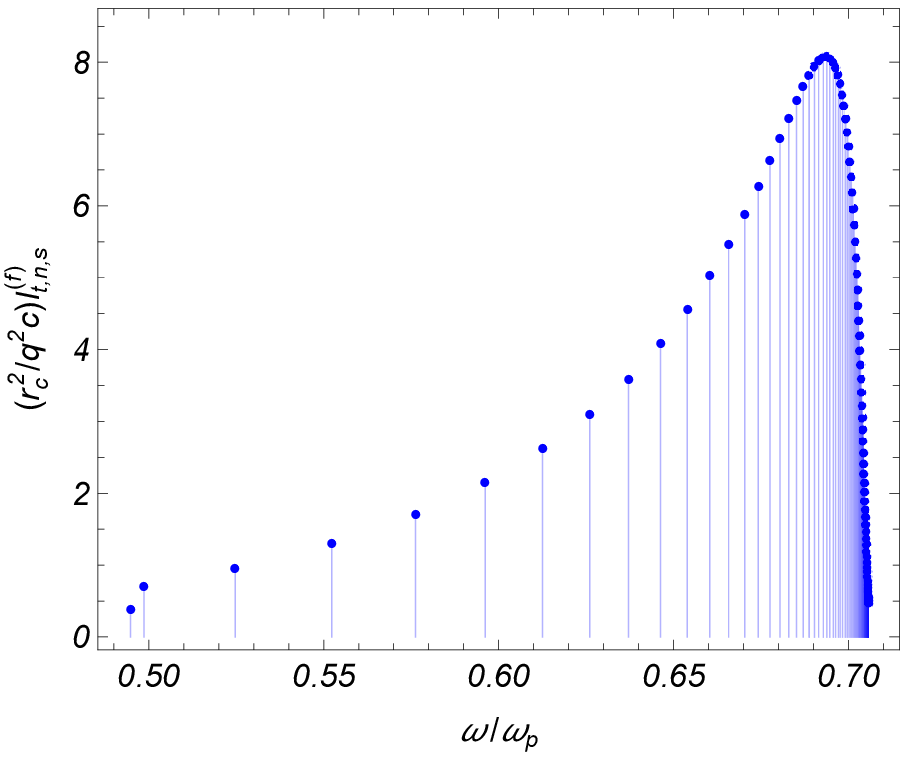,width=7.5cm,height=6.cm} & \quad %
\epsfig{figure=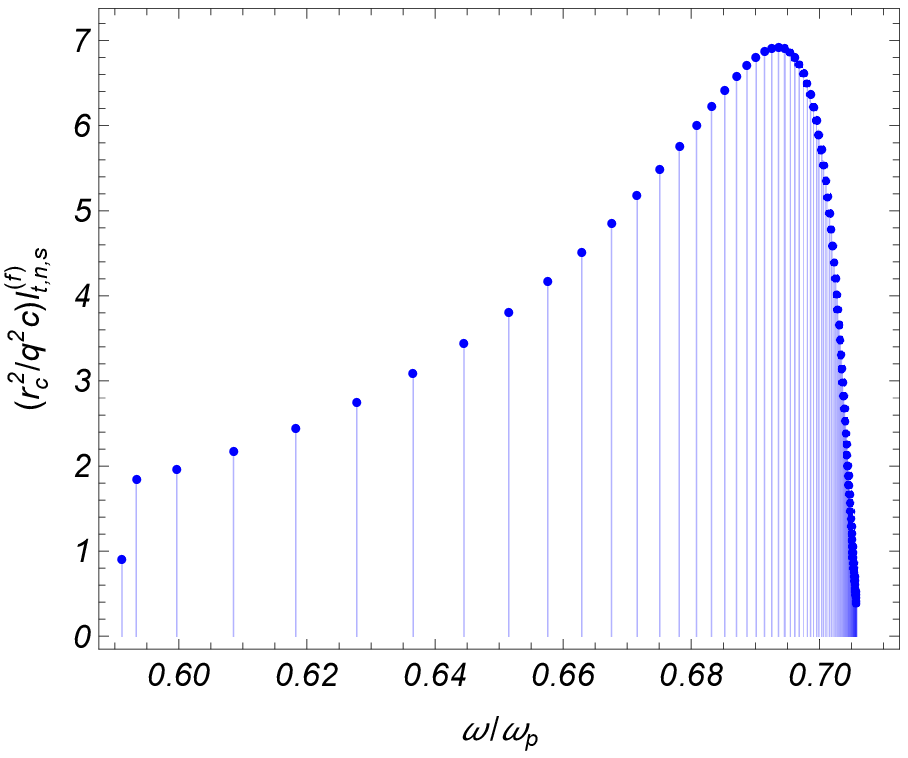,width=7.5cm,height=6.cm}%
\end{tabular}%
\end{center}
\caption{Total energy flux versus the eigenfrequencies of the surface
polariton modes in the model with (\protect\ref{eps1}) and for $\protect%
\varepsilon _{0}=1$ for the values of the parameters $r_{0}/r_{c}=0.95$, $%
r_{c}\protect\omega _{p}/c=10$, and $\protect\beta =0.9,0.75$ for the left
and right panels.}
\label{fig9}
\end{figure}

\begin{figure}[tbph]
\begin{center}
\begin{tabular}{cc}
\epsfig{figure=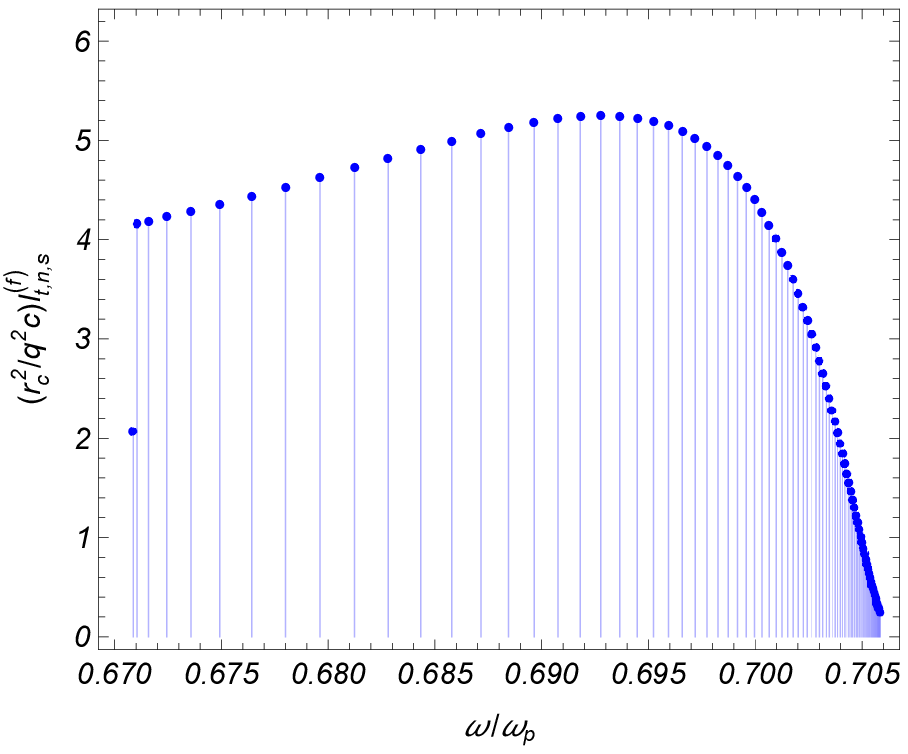,width=7.5cm,height=6.cm} & \quad %
\epsfig{figure=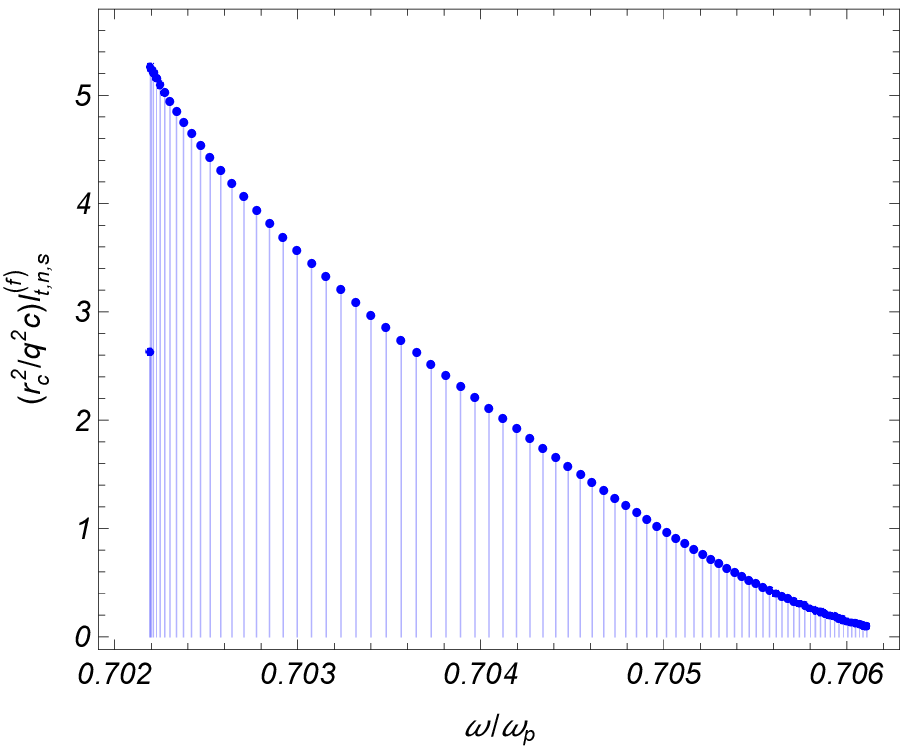,width=7.5cm,height=6.cm}%
\end{tabular}%
\end{center}
\caption{The same as in Figure \protect\ref{fig9} for $\protect\beta =0.5$
(left panel) and $\protect\beta =0.25$ (right panel).}
\label{fig10}
\end{figure}

In Figures \ref{fig9} and \ref{fig10}, the total energy flux is displayed as
a function of the eigenfrequencies $\omega _{n}/\omega _{p}$ of the surface
polaritonic modes. As for the example in Figure \ref{fig8}, we have taken $%
\varepsilon _{0}=1$, $\varepsilon _{1}=1-\omega _{p}^{2}/\omega ^{2}$, and $%
0\leq n\leq 80$. The numerical evaluation is done for $r_{0}/r_{c}=0.95$, $%
r_{c}\omega _{p}/c=10$. The left and right panels of Figure \ref{fig9}
correspond to $\beta _{0}=0.9$ and $\beta _{0}=0.75$. Correspondingly, the
left and right panel of Figure \ref{fig10} are plotted for $\beta _{0}=0.5$
and $\beta _{0}=0.25$. Note that, in order to show the dependence on the
charge velocity, in Figures \ref{fig9} and \ref{fig10} we have presented the
quantity $(r_{c}^{2}q^{2}/c)I_{\mathrm{t},n,s}^{(f)}$ instead of $%
(r_{c}^{2}q^{2}/v)I_{\mathrm{t},n,s}^{(f)}$ in Figure \ref{fig8}. 

The numerical examples, we have discussed above for the properties of the
roots and for the energy fluxes, are presented in terms of scale invariant
combinations of the parameters. This allows to specify the corresponding
results for different values of the cylinder radius. As already stated, for
given $\beta _{0}$ and $\varepsilon _{1}/\varepsilon _{0}$ the SP eigenmodes
with respect to $u=k_{z}r_{c}$ do not depend on the radius of the cylinder.
The radiation wavelength, $\lambda _{n,s}=2\pi r_{c}/u_{n,s}$, is controlled
by the choice of the waveguide radius. The recent advances in
nanofabrication allow to design cylindrical waveguides with radii in a
sufficiently wide range, from millimeters to nanometers (see, e.g., Refs. 
\cite{Abaj10,Atak13,Isla20}). One can control the wavelength of radiated SPs
by an appropriate choice of the waveguide radius and negative-permittivity
medium. Materials and artificially constructed subwavelength structures are
available with plasma frequency in the visible, infrared and terahertz
frequency ranges. The electron beam in TEMs provides an example of high
quality source in relatively wide energy range from 50 keV to 500 keV. The
same beam from TEMs can be used to drill nanometre-scale cylindrical holes
in a medium.

Our main concern in the discussion above was the radiation for a single
charge. Based on the results obtained, we can investigate the radiation from
a bunch of $N$ particles with velocities $\mathbf{v}_{m}$, $m=1,2,\ldots ,N$%
, parallel to the cylinder axis. For the current density one has%
\begin{equation}
j_{l}(x)=\delta _{3l}q\sum_{m=1}^{N}\frac{v_{m}}{r_{m}}\delta (\mathbf{r}-%
\mathbf{r}_{0m}(t)),  \label{jlb}
\end{equation}%
where $\mathbf{r}_{0m}(t)=(r_{m},\phi _{m},z_{m}+v_{m}t)$ in cylindrical
coordinates. Assuming that for all the particles $r_{m}<r_{c}$, the
electromagnetic fields corresponding to the radiated SPs are obtained by
summing the fields for separate charges. For example, the formula for the
components of the vector potential reads%
\begin{eqnarray}
A_{l}^{\mathrm{(P)}}(x) &=&\frac{2q}{cr_{c}}\sum_{m=1}^{N}v_{m}%
\sideset{}{'}{\sum}_{n=0}^{\infty }\sum_{s}Q_{m,n}(u)\sum_{p=\pm 1}\frac{%
R_{m,n+p}(u,r/r_{c})}{p^{l-1}uW_{m,n+p}^{I}}  \notag \\
&&\times \cos \left( u\xi _{m}/r_{c}\right) \sin \left[ l\pi /2-n\left( \phi
-\phi _{m}\right) \right] |_{u=u_{m,n,s}},  \label{Alb}
\end{eqnarray}%
where $\xi _{m}=v_{m}t-z+z_{m}$ and the expressions for $R_{m,n}(u,r/r_{c})$%
, $W_{m,n}^{I}$, $Q_{m,n}(u)$ are obtained from the corresponding
expressions for the functions without the index $m$ by the replacements $%
\gamma _{j}\rightarrow \gamma _{m,j}=\sqrt{1-v_{m}^{2}\varepsilon _{j}/c^{2}}
$ and $r_{0}\rightarrow r_{m}$. For monoenergetic bunches with transverse
beam size smaller than the radiation wavelength (the specific condition will
depend on the energy of the beam as well) we can approximate the general
formula (\ref{Alb}) taking $v_{m}=v$, $r_{m}=r_{0}$, $\phi _{m}=0$. In this
simple case the expressions for the radiation fields are obtained from the
formulas given in Section \ref{sec:AHE} making the replacements $\cos \left(
u\xi /r_{c}\right) \rightarrow \sum_{m}\cos \left( u\xi _{m}/r_{c}\right) $
and $\sin \left( u\xi /r_{c}\right) \rightarrow \sum_{m}\sin \left( u\xi
_{m}/r_{c}\right) $. For the energy flux through the plane $z=\mathrm{const}$
we get an expression which is obtained from (\ref{Iksi}) replacing the
product of sin functions by $\sum_{m,m^{\prime }}\sin \left( u_{n,s}\xi
_{m}/r_{c}\right) \sin \left( u_{n,s^{\prime }}\xi _{m^{\prime
}}/r_{c}\right) $. By using the same averaging procedure, we can see that
the energy fluxes for SPs radiated by a bunch with longitudinal distribution
function $f(z)$ are obtained from (\ref{Ifi}) and (\ref{Ife}) adding the
factor $N[1+(N-1)\left\vert g(u/r_{c})\right\vert ^{2}]$ in the right-hand
sides, where $g(w)=\int_{-\infty }^{+\infty }dz\,\,e^{-iwz}f(z)$ is the
longitudinal bunch form factor. The second term in the square brackets
describes the coherent effects in the radiation of SPs. Note that the
longitudinal form factor depends on the frequency and on the monoenergetic
bunch velocity in the form of the ratio $\omega /v$. This property is a
direct consequence of homogeneity of the problem under consideration along
the $z$-direction. A similar longitudinal form factor appears also for other
types of radiation processes, such as Cherenkov radiation, Smith-Purcell
radiation, etc. (see, e.g., \cite{Shib98}-\cite{Tade22}). The coherence
effects have been used to increase significantly the radiation intensity in
different spectral ranges and also in beam diagnostics. For the radiation
wavelengths of the order of transverse beam size or smaller the effect of
the transverse form factor on the coherence properties becomes significant.
In particular, the dependence on the energy of the beam is more pronounced.

In the discussion above we have considered an idealized problem where the
dielectric permittivities of the media inside and outside the cylinder are
taken to be real. A small imaginary part of the permittivity of the exterior
medium was introduced in Section \ref{sec:GF} in order to specify the
contour of the integration over $k_{z}$ near the poles, corresponding to the
roots of the eigenmode equation for SPs. The approach we have
described can be considered as a first step to the investigation of the
surface polariton generation in more realistic setups with energy losses.
The damping of SPs arising from the imaginary part of
dielectric permittivity medium is one of the main limitations for practical
applications in plasmonic devices. The energy dissipation, primarily in the
form of Ohmic losses, limits the energy accumulated by SPs
and may significantly reduce their propagation distances (for various decay
channels of SPs energy dissipation see, e.g., \cite{Bori17}).
In particular, that is the case for the commonly used plasmonic materials in
the optical range such as silver and gold. Related to this, the development
of various approaches and mechanisms aiming to reduce or compensate the
energy losses remains among the main directions in plasmonics. They can be
categorized into three main groups \cite{Bori17,Bolt11,Khur12}. The first
one is the choice of suitable material for negative-permittivity medium. The
list of low loss plasmonic materials in mid infrared and terahertz spectral
ranges include various kinds of doped semiconductors, superconductors,
transparent conducting oxides, different types of metamaterials, topological
insulators and 2D Dirac materials like graphene (see, for example, Refs. 
\cite{Chan11}-\cite{Qin23} and references therein). An important advantage
with these classes of plasmonic materials is the possibility to actively
tune the plasma frequency. For example, that can be done by the choice of
doping level in doped semiconductors and by electrostatic gating in
graphene. The second direction to reduce the dissipative losses of surface
plasmons corresponds to the engineering the shape and size of the structure
along which the waves propagate. They include grating type structures with
different geometries and metamaterials with controllable electromagnetic
characteristics. And finally, the third direction of investigations uses
gain media to compensate the energy losses of SPs.

Note that the expressions (\ref{Gi}) and (\ref{Ge}) are valid for general
case of complex dielectric functions $\varepsilon _{0}$ and $\varepsilon
_{1} $, and they can be used for the evaluation of the electromagnetic
fields in the problem at hand without specifying those functions. The scheme
is similar to that we have described for the evaluation of the SP contributions: first we evaluate the vector potential by using
Eq. (\ref{Aix}) and then the scalar potential, electric and magnetic fields
by standard formulas in classical electrodynamics. In the next section that
procedure is described for the axial component of the electric field which
determines the total energy losses of the charged particle.

\section{Energy losses}

\label{sec:EnLoss}

In the discussion of the properties of the radiated SPs we have considered
an idealized case where the imaginary part of the dielectric functions were
ignored. We can investigate the total energy losses by a charged particle
for general case of dielectric permittivities by using the expressions (\ref%
{Gexp}), (\ref{Gi}) for the components of the Green tensor. Those
expressions are also valid for dielectric functions having imaginary parts.
Denoting by $\mathbf{E}(x)$ the electric field generated by the charge at
the spacetime point $x=(t,\mathbf{r})$, the energy loss per unit length
along the trajectory of the charge (the work of the field on the charge) is
expressed as%
\begin{equation}
\frac{dW}{dz}=qE_{3}(x)|_{\mathbf{r}\rightarrow \mathbf{r}_{0}(t)}.
\label{dW}
\end{equation}%
By making use of the Fourier expansion for the axial component of the
electric field,%
\begin{equation}
E_{3}(x)=\sum_{n=-\infty }^{+\infty }\int_{-\infty }^{+\infty
}dk_{z}\,E_{3,n}(k_{z},r)e^{ik_{z}(z-vt)+in\phi },  \label{Efour}
\end{equation}%
and the properties of the Fourier component $E_{3,n}(k_{z},r)$, the formula
is rewritten as%
\begin{equation}
\frac{dW}{dz}=2q\lim_{r\rightarrow r_{0}}\sum_{n=-\infty }^{+\infty }\mathrm{%
Re}\left[ \int_{0}^{\infty }dk_{z}\,E_{3,n}(k_{z},r)\right] .  \label{dW1}
\end{equation}%
The expression for $E_{3,n}(k_{z},r)$ can be found based on the
representations (\ref{Gi}), by the scheme similar to that we have used in
Section \ref{sec:AHE} for the contributions of SPs in the case of real
dielectric functions $\varepsilon _{0}$ and $\varepsilon _{1}$.

In this way, the energy losses are presented in the form%
\begin{equation}
\frac{dW}{dz}=\frac{dW^{(0)}}{dz}-\frac{4q^{2}}{\pi r_{c}^{2}}\mathrm{Im}%
\left\{ \sum_{n=0}^{\infty }\delta _{n}\int_{0}^{\infty }du\,\frac{u\gamma
_{0}^{2}}{\varepsilon _{0}W_{n}^{I}}\left[ W_{n}^{K}+\frac{K_{n}(u\gamma
_{1})}{2\gamma _{0}u\alpha _{n}(u)}\sum_{p=\pm 1}\frac{K_{n+p}(u\gamma _{1})%
}{W_{n+p}^{I}}\right] I_{n}^{2}(u\gamma _{0}r_{0}/r_{c})\right\} ,
\label{dW2}
\end{equation}%
where%
\begin{equation}
\frac{dW^{(0)}}{dz}=\frac{2q^{2}}{\pi }\lim_{r\rightarrow r_{0}}\mathrm{Im}%
\left[ \sum_{n=-\infty }^{\infty }\int_{0}^{\infty }dk_{z}\,\frac{k_{z}}{%
\varepsilon _{0}}\gamma _{0}^{2}I_{n}(\gamma _{0}k_{z}r_{<})K_{n}(\gamma
_{0}k_{z}r_{>})\right] ,  \label{dW0}
\end{equation}%
with $r_{>}=\mathrm{max}(r_{0},r)$ and $r_{<}=\mathrm{min}(r_{0},r)$. In the
second term of (\ref{dW2}) we have passed to the integration over $u$ in
accordance with (\ref{u}) and the notation%
\begin{equation}
W_{n}^{K}=\gamma _{0}K_{n}(\gamma _{1}u)K_{n+1}(\gamma _{0}u)-\gamma
_{1}K_{n}(\gamma _{0}u)K_{n+1}(\gamma _{1}u),  \label{WnK}
\end{equation}%
is introduced. Other notations are the same as those used in the
consideration above. However, now $\varepsilon _{0}$ and $\varepsilon _{1}$,
in general, are complex functions and, hence, the same is the case for $%
\gamma _{0}$ and $\gamma _{1}$, defined in (\ref{gamj}). The contribution (%
\ref{dW0}) does not depend on the cylinder radius $r_{c}$ and it corresponds
to the energy losses in a homogeneous medium with dielectric permittivity $%
\varepsilon _{0}$ (bulk losses). These energy losses have been extensively
investigated in the literature both theoretically and experimentally. Here
we note that most of the previous studies consider the spectral density of
the energy loss probability per unit length, $dP\left( \omega \right) /dz$,
with the relation $dW/dz=\int_{0}^{\infty }d\omega \,\omega dP\left( \omega
\right) /dz$. The series over $n$ in (\ref{dW0}) is summed by using the
formula from \cite{Prud2} and we get%
\begin{equation}
\frac{dW^{(0)}}{dz}=\frac{2q^{2}}{\pi }\lim_{r\rightarrow r_{0}}\mathrm{Im}%
\left[ \int_{0}^{\infty }dk_{z}\,\frac{k_{z}}{\varepsilon _{0}}\gamma
_{0}^{2}K_{0}(k_{z}|r-r_{0}|\sqrt{1-\beta _{0}^{2}})\right] .  \label{dW0b}
\end{equation}%
For a transparent medium ($\varepsilon _{0}$ is real) and under the
condition $\beta _{0}^{2}<1$, the integrand is real and $dW^{(0)}/dz=0$. For
transparent medium and under the Cherenkov condition $\beta _{0}^{2}>1$ the
imaginary part of the Macdonald function in (\ref{dW0b}) is expressed in
terms of the Bessel function as $\pi J_{0}(k_{z}|r-r_{0}|\sqrt{\beta
_{0}^{2}-1})/2$. In this case the limit $r\rightarrow r_{0}$ can be taken
directly in the integrand and from (\ref{dW0b}) we get the standard
expression for the radiation intensity of the Cherenkov radiation in a
homogeneous medium.

The second term on the right-hand side of (\ref{dW2}) is induced by the
difference of the dielectric permittivity in the region $r>r_{c}$ from $%
\varepsilon _{0}$. By using the definitions for $W_{n}^{I}$, $W_{n}^{K}$,
and $\alpha _{n}(u)$, the corresponding expression is written in more
explicit form%
\begin{eqnarray}
\frac{dW}{dz} &=&\frac{dW^{(0)}}{dz}-\frac{4q^{2}}{\pi r_{c}^{2}}\mathrm{Im}%
\left\{ \sum_{n=0}^{\infty }\delta _{n}\int_{0}^{\infty }du\,\frac{u}{%
\varepsilon _{0}}\gamma _{0}^{2}I_{n}^{2}(u\gamma _{0}r_{0}/r_{c})\frac{%
K_{n}(\gamma _{0}u)}{I_{n}(\gamma _{0}u)}\right.  \notag \\
&&\times \left. \frac{\left[ \gamma _{1}\frac{I_{n}^{\prime }(\gamma _{0}u)}{%
I_{n}(\gamma _{0}u)}-\gamma _{0}\frac{K_{n}^{\prime }(\gamma _{1}u)}{%
K_{n}(\gamma _{1}u)}\right] \left[ \varepsilon _{0}\gamma _{1}\frac{%
K_{n}^{\prime }\left( \gamma _{0}u\right) }{K_{n}\left( \gamma _{0}u\right) }%
-\varepsilon _{1}\gamma _{0}\frac{K_{n}^{\prime }\left( \gamma _{1}u\right) 
}{K_{n}\left( \gamma _{1}u\right) }\right] -\left( \frac{n\beta }{u}\frac{%
\varepsilon _{0}-\varepsilon _{1}}{\gamma _{0}\gamma _{1}}\right) ^{2}}{%
\left[ \gamma _{1}\frac{I_{n}^{\prime }(\gamma _{0}u)}{I_{n}(\gamma _{0}u)}%
-\gamma _{0}\frac{K_{n}^{\prime }(\gamma _{1}u)}{K_{n}(\gamma _{1}u)}\right] %
\left[ \varepsilon _{0}\gamma _{1}\frac{I_{n}^{\prime }\left( \gamma
_{0}u\right) }{I_{n}\left( \gamma _{0}u\right) }-\varepsilon _{1}\gamma _{0}%
\frac{K_{n}^{\prime }\left( \gamma _{1}u\right) }{K_{n}\left( \gamma
_{1}u\right) }\right] -\left( \frac{n\beta }{u}\frac{\varepsilon
_{0}-\varepsilon _{1}}{\gamma _{0}\gamma _{1}}\right) ^{2}}\right\} .
\label{dW3}
\end{eqnarray}%
This expression coincides with that obtained from the energy loss
probability found in \cite{Wals91}. Note that the zeros of the denominator
determine the SP eigenmodes (compare with (\ref{EigEq})). In the special
case of the axial motion with $r_{0}=0$ the only nonzero contribution comes
from the mode $n=0$ and one gets%
\begin{equation}
\frac{dW}{dz}=\frac{dW^{(0)}}{dz}+\frac{2q^{2}}{\pi r_{c}^{2}}\mathrm{Im}%
\left[ \int_{0}^{\infty }du\,\frac{u}{\varepsilon _{0}}\gamma _{0}^{2}\frac{%
\varepsilon _{0}\gamma _{1}K_{0}\left( \gamma _{1}u\right) K_{1}\left(
\gamma _{0}u\right) -\varepsilon _{1}\gamma _{0}K_{1}\left( \gamma
_{1}u\right) K_{0}\left( \gamma _{0}u\right) }{\varepsilon _{0}\gamma
_{1}K_{0}\left( \gamma _{1}u\right) I_{1}\left( \gamma _{0}u\right)
+\varepsilon _{1}\gamma _{0}K_{1}\left( \gamma _{1}u\right) I_{0}\left(
\gamma _{0}u\right) }\right] .  \label{dWaxial}
\end{equation}%
This result was obtained in \cite{Bolo62,Zutt86}.

Another special case corresponds to the non-relativistic limit, $\beta \ll 1$%
. Assuming that $\beta ^{2}|\varepsilon _{j}|\ll 1$, $j=0,1$, to the leading
order we can put $\gamma _{j}=1$. In the same order, one gets $%
W_{n}^{I}\approx -1/u$, $W_{n}^{K}\approx 0$, and the function $\alpha
_{n}(u)$ is approximated by (\ref{alfnr}). From (\ref{dW3}), for the leading
order contribution to the energy losses we find%
\begin{equation}
\frac{dW}{dz}\approx \frac{dW^{(0)}}{dz}-\frac{4q^{2}}{\pi r_{c}^{2}}%
\sum_{n=0}^{\infty }\delta _{n}\mathrm{Im}\left[ \int_{0}^{\infty }du\,\frac{%
u}{\varepsilon _{0}}\frac{uK_{n}(u)K_{n}^{\prime }(u)I_{n}^{2}(ur_{0}/r_{c})%
}{\frac{\varepsilon _{0}}{\varepsilon _{0}-\varepsilon _{1}}%
+uI_{n}(u)K_{n}^{\prime }(u)}\right] .  \label{dWnr}
\end{equation}%
The corresponding result for the energy loss probability has been widely
discussed in the literature (see \cite%
{Riva00,Zaba89,Wals89,Riva95,Aris01,Gerv03,Pita07}).

\begin{figure}[tbph]
\begin{center}
\epsfig{figure=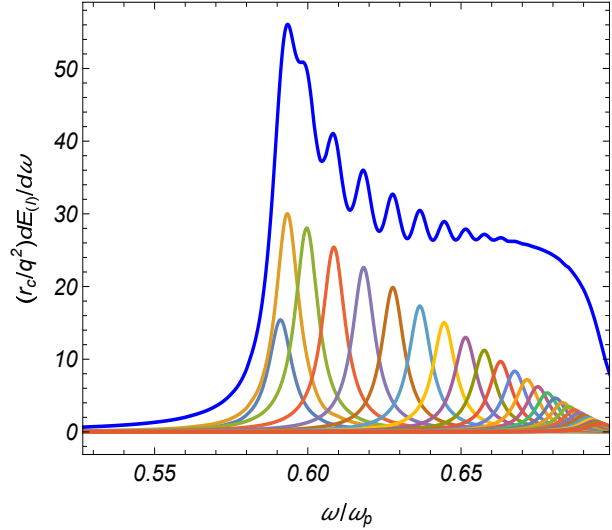,width=8.cm,height=7.cm}
\end{center}
\caption{The spectral density of the energy loss per unit time as a function
of frequency. The graphs are plotted for $\protect\varepsilon _{0}=1$ and
for the dispersion of the function $\protect\varepsilon _{1}(\protect\omega %
) $ described by (\protect\ref{eps1}) The values of the parameters are given
in the text. The contributions of the modes with fixed $n$, $0\leq n\leq 25$%
, are plotted as well.}
\label{fig11}
\end{figure}

For the numerical example of the energy losses we have considered the case
where $\varepsilon _{0}=1$ and the dielectric function for the medium in the
region $r>r_{c}$ is described by (\ref{eps1}). In this special case one has $%
dW^{(0)}/dz=0$. Let us introduce the spectral density of the energy loss per
unit time, $dE_{\mathrm{(l)}}(\omega )/d\omega $, in accordance with%
\begin{equation}
\frac{dW}{dz}=-\frac{1}{v}\int_{0}^{\infty }d\omega \,\frac{dE_{\mathrm{(l)}}%
}{d\omega }.  \label{dEl}
\end{equation}%
By using (\ref{dW2}) we get%
\begin{equation}
\frac{dE_{\mathrm{(l)}}}{d\omega }=\frac{4q^{2}}{\pi r_{c}\gamma ^{2}}%
\sideset{}{'}{\sum}_{n=0}^{\infty }\mathrm{Im}\,\left[ u\frac{W_{n}^{K}}{%
W_{n}^{I}}+\frac{\gamma K_{n}(u\gamma _{1})}{2\alpha _{n}(u)}\sum_{p=\pm 1}%
\frac{K_{n+p}(u\gamma _{1})}{W_{n}^{I}W_{n+p}^{I}}\right] I_{n}^{2}\left( 
\frac{ur_{0}}{\gamma r_{c}}\right) ,  \label{dEl1}
\end{equation}%
where $\gamma =1/\sqrt{1-\beta ^{2}}$ is the relativistic factor and $%
u=r_{c}\omega /v$. Now, in the definitions of the functions $W_{n}^{I}$, $%
W_{n}^{K}$, and $\alpha _{n}(u)$ one has $\gamma _{0}=1/\gamma $. In Figure %
\ref{fig11} the spectral density of the energy loss $dE_{\mathrm{(l)}%
}(\omega )/d\omega $ is presented in units of $q^{2}/r_{c}$ versus the ratio 
$\omega /\omega _{p}$. The graphs are plotted for $\eta /\omega _{p}=10^{-2}$%
, $r_{0}/r_{c}=0.95$, $\beta =0.75$, and $r_{c}\omega _{p}/c=10$. We have
also displayed the separate contributions of the modes with different $n$, $%
dE_{\mathrm{(l)}n}(\omega )/d\omega $, $0\leq n\leq 25$, defined as $dE_{%
\mathrm{(l)}}(\omega )/d\omega =\sum_{n=0}^{\infty }dE_{\mathrm{(l)}%
n}(\omega )/d\omega $. For $n\geq 1$ the frequency corresponding to the
maximum of $dE_{\mathrm{(l)}n}(\omega )/d\omega $ increases with increasing $%
n$ and the maximal value of that quantity decreases with increasing $n$. The
curve with the minimal value for $\omega /\omega _{p}$ at the peak
correspond to the mode $n=0$.

\section{Conclusion}

\label{sec:Conc}

We have investigated the radiation emitted by a charge uniformly moving
inside a dielectric cylinder, parallel to its axis, assuming that the
cylinder is loaded in a homogeneous medium. For evaluation of the
electromagnetic fields generated inside and outside the cylinder the Green
tensor from \cite{Grig95} has been used. The corresponding expressions allow
to study both the cases of the medium with negative dielectric permittivity
in the spectral range under consideration inside and outside the cylinder.
We have specified the investigation for the second case that will include
the possibility of the charge motion in the vacuum. The required components
of the Green tensor Fourier image are expressed as (\ref{Gi}) and (\ref{Ge}%
). Neglecting the imaginary part of dielectric permittivity, the Fourier
components have poles corresponding to SPs. The respective contributions to
the Green tensor are separated and they have been used in evaluating the
field potentials and strengths inside and outside the cylinder. In general,
both the transversal and longitudinal components of the electric and
magnetic fields for excited SPs differ from zero. The fields exponentially
decay in the exterior medium and they are mainly confined in the region of
the thickness of the order $\lambda _{\mathrm{sp}}/(2\pi \gamma _{1})$ near
the cylinder surface. The localization radius decreases with increasing
velocity of the charge and it can be essentially smaller compared with the
radiation wavelength $\lambda _{\mathrm{sp}}$. The fields are expressed in
terms of the eigenvalues for the projection of the wave vector along the
cylinder axis and we have discussed their distribution as functions of the
parameters and in the asymptotic regions. In particular, in the
non-relativistic limit the SP modes are present in the region $-1\leq
\varepsilon _{1}/\varepsilon _{0}<0$ for the ratio of the dielectric
functions. The relativistic effects may essentially enlarge the region for $%
\varepsilon _{1}/\varepsilon _{0}$ allowing the existence of the SP modes.
We have specified the general consideration for the case of Drude dispersion
in the exterior medium. The impact parameter $r_{0}$ enters in the
expressions for the fields through the function $I_{n}(\gamma
_{0}ur_{0}/r_{c})$ in the definition (\ref{Qn}) and, for a given frequency,
the absolute values for the components of the fields monotonically increase
with increasing $r_{0}$. The general formulas are essentially simplified in
the special case of axial motion when the only nonzero contribution to the
radiation fields come from the mode with $n=0$. In this special case the
magnetic field is transversal and the electric and magnetic fields are
orthogonal.

Having the electric and magnetic fields for SPs, in Section \ref{sec:Flux}
we have evaluated the corresponding mean energy fluxes in the exterior and
interior regions, given by (\ref{Ifi}) and (\ref{Ife}). The exterior energy
flux, corresponding to the negative-permittivity medium, is negative (flux
along the direction opposite to the charge motion), whereas the flux inside
the cylinder (positive-permittivity medium) is positive (directed along the
direction of the charge motion). The total flux is dominated by the interior
contribution and it is positive. In the non-relativistic limit the energy
fluxes are proportional to the charge velocity. The relativistic effects may
essentially increase the radiated energy. Other important features of
relativism include the narrowing the confinement region of the SP fields
near the cylinder surface in the exterior region, enlarging the frequency
range for radiated SPs, and the decrease of the cotuff factor for radiation
at small wavelengths compared with the cylinder radius. The energy fluxes at
those wavelengths are approximated by (\ref{Ifias}) and (\ref{Ifeas}).
Relatively simple expressions for interior and exterior energy fluxes, (\ref%
{Ii0}) and (\ref{Ie0}), are obtained in the special case of the axial
motion. The features clarified by asymptotic analysis of exact formulas are
confirmed by numerical data. We have presented the latter in terms of
dimensionless combinations of the parameters that allows to specify the
results for different values of the waveguide radius and for different
spectral ranges. Given the radiation fields generated by a single charge,
the generalization is straightforward for a bunch of particles moving
parallel to the axis of the cylinder. For example, the corresponding vector
potential is expressed as (\ref{Alb}). In the special case of a
monoenergetic bunch with transverse size smaller than the radiation
wavelength, the collective effects in the energy fluxes on a given frequency
appear through the bunch longitudinal from factor.

By using the expressions for the components of the Green tensor, we have
also considered the total energy losses for general case of dielectric
functions of the exterior and interior media with imaginary parts. The
general formula is given by (\ref{dW2}) or, equivalently, by (\ref{dW3}).
The latter coincides with the result obtained from the energy loss
probability previously considered in the literature and includes various
special cases widely discussed before. Similar to the case of the SP energy
fluxes, the numerical analysis is provided in scale invariant form that
allows to specify the result for special cases of the parameters (e.g.,
cylinder radius and plasma frequency for the negative-permittivity medium).

In our consideration the exterior medium occupies the region $r_{c}<r<\infty 
$. Based on the features described above, we expect that the obtained
expressions of the SP energy fluxes for a given wavelength will approximate
the corresponding results for the medium with finite extension, $r_{c}<r<r_{%
\mathrm{ext}}$, if the thickness of the cylindrical layer $r_{\mathrm{ext}%
}-r_{c}$ is larger than the confinement radius for the SPs on that
wavelength. Note that the Green tensor in the problem with finite exterior
layer can be found based on the recurrence procedure developed in \cite%
{Grig95} for general number of coaxial cylindrical \ layers. Another
application of the results presented in this paper could be the
investigation of the transversal forces acting on the charge in the case of
paraxial motion. Those forces are of interest in studies of beam stabilities
in particle accelerators. And finally, the problem we have considered is
exactly solvable within the framework of classical electrodynamics and the
corresponding results may serve as a tool to verify the accuracy of various
approximate methods and simulations used for the investigation of surface
polaritons in more complicated geometries of interfaces.

\section*{Acknowledgement}

A.A.S. was supported by the Science Committee of RA, in the frames of the
research project No. 21AG-1C047. L.Sh.G. and H.F.K. were supported by the
Science Committee of RA, in the frames of the research project No.
21AG-1C069.

\end{document}